\documentclass[12pt,letterpaper]{article}
\usepackage{booktabs,multirow}  
\usepackage{graphicx,lscape}
\usepackage{amsmath,amssymb}
\usepackage{mathtools}
\usepackage{bm}
\usepackage{threeparttable}
\usepackage{url}
\usepackage{graphicx}
\usepackage{caption, makecell}
\usepackage[dvipsnames]{xcolor}
\usepackage{yfonts}
\usepackage{natbib}
\usepackage{relsize}
\usepackage{rotating}
\usepackage{chngcntr}

\begin{document}
\baselineskip=24pt

\begin{center}
\Large \bf
Adjusting for publication bias in meta-analysis via inverse probability weighting using clinical trial registries\\
\end{center}

\vspace{3mm}

\begin{center}
Ao Huang \\
\sl { Department of Biomedical Statistics, Graduate School of Medicine, \\
Osaka University, 2-2, Yamadaoka, Suita, Osaka 565-0871, Japan.\\
Email: huangao@biostat.med.osaka-u.ac.jp}\\
\vspace{6mm}
Kosuke Morikawa \\
{\it Graduate School of Engineering Science,
Osaka University \\ 
Toyonaka, Osaka 560-8531, Japan \\
E-mail: morikawa@sigmath.es.osaka-u.ac.jp
} \\ 
\vspace{6mm}
Tim Friede  \\
{\it Department of Medical Statistics, University Medical Center G\"{o}ttingen, \\
Humboldtallee 32 G\"{o}ttingen, Germany 37073 \\
E-mail: tim.friede@med.uni-goettingen.de
} \\
\vspace{6mm}
Satoshi Hattori \\
{\it Department of Biomedical Statistics, Graduate School of Medicine, and \\
Integrated Frontier Research for Medical Science Division, Institute for Open and Transdisciplinary ResearchInitiatives (OTRI), Osaka University, \\
Yamadaoka 2-2, Suita City, Osaka 565-0871, Japan \\
E-mail: hattoris@biostat.med.osaka-u.ac.jp
} \\ 
\end{center}

\begin{center}
Running title: IPW for publication bias adjustment
\end{center}

\clearpage

\begin{abstract}
Publication bias is a major concern in conducting systematic reviews and meta-analyses. Various sensitivity analysis or bias-correction methods have been developed based on selection models and they have some advantages over the widely used bias-correction method of the trim-and-fill method. However, likelihood methods based on selection models may have difficulty in obtaining precise estimates and reasonable confidence intervals or require a complicated sensitivity analysis process. In this paper, we develop a simple publication bias adjustment method utilizing information on conducted but still unpublished trials from clinical trial registries. We introduce an estimating equation for parameter estimation in the selection function by regarding the publication bias issue as a missing data problem under missing not at random. With the estimated selection function, we introduce the inverse probability weighting (IPW) method to estimate the overall mean across studies. Furthermore, the IPW versions of heterogeneity measures such as the between-study variance and the $I^2$ measure are proposed. We propose methods to construct asymptotic confidence intervals and suggest intervals based on parametric bootstrapping as an alternative. Through numerical experiments, we observed that the estimators successfully eliminate biases and the confidence intervals had empirical coverage probabilities close to the nominal level. On the other hand, the asymptotic confidence interval is much wider in some scenarios than the bootstrap confidence interval. Therefore, the latter is recommended for practical use. 
\\ 

Key words: Clinical trial registry; Missing not at random; Propensity score; Sensitivity analysis; Systematic review 
\end{abstract}

\clearpage

\section{Introduction}
Meta-analyses play a very important role in medical research and may have substantial impact in establishing sound medial evidence. Meta-analysts try to gather all the available evidences by conducting systematic literature searches including not only the scientific literature but also the so-called grey literature such as documents for regulation of new drug applications and conference abstracts \citep{gopalakrishnan2013}.  Despite of such pain-taking efforts, it is very hard to collect all information; then the reporting biases may arise when some negative results might not be reported by investigators or are not likely to be accepted by scientific journals or might be presented in a way that they become positive. Especially when it comes to the situation that publication status (publication or non-publication) depends on the nature and the direction of research findings, it was usually referred to as the publication bias \citep{thornton2000}. 

The funnel plot and the trim-and-fill method are among the most widely used methods to identify and adjust for publication bias \citep{egger1997, duval2000}. Despite of their simple interpretability through graphical presentation, results obtained by these methods may be misleading \citep{terrin2003, peters2007}. Modeling the selective publication process by a selection model may yield more reliable and interpretable results to quantify the impact of publication bias \citep{carpenter2009, schwarzer2010}. The Copas-Shi selection model was suggested to be preferable to the trim-and-fill method by \citet{schwarzer2010}. It was an adoption of the Heckman selection model, which was first proposed in the context of econometrics, then introduced to the area of meta-analysis by \citet{copas1999} and \citet{copas2000}. A notable feature of the Copas-Shi selection model is that it modeled the selection process based on a simple Gaussian latent variable, which can be easily linked to any normally distributed population model for its mathematical nature. This simplicity led wide extensions to more complicated meta-analyses such as the network meta-analysis \citep{mavridis2013} and the diagnostic meta-analysis \citep{hattori2018,piao2019,li2021}, interpretation of the Heckman-type selection function might not be satisfactory in medical research. Selection functions defined with the test statistics used in each publication might be more appealing since \emph{P}-values might be a very influential factor for the decision to publish. \citet{preston2004} discussed maximum conditional likelihood estimation with a series of one-parameter selection functions based on the empirical $P$-values; \citet{copas2013} proposed a likelihood-based sensitivity analysis method with the selection function modeling the Wald-type statistics directly. Following \citet{copas2013}, we denote these selection functions as $t$-type selection functions. Since inference of these methods is based on published data only, the maximization of the conditional likelihood can be computationally challenging even only with one parameter, hence a sensitivity analysis is recommended in practice by both \citet{preston2004} and \citet{copas2013}. With some sensitivity parameters fixed in a plausible range, then the impact of the publication bias can be studied. Indeed, as will be demonstrated in our simulation study, the maximum likelihood estimation conditional on published might be hard to get converged and result in an unreasonable confidence interval. 

Registration of study protocols in clinical trial registries is a non-statistical approach against selective publication; by prospectively registering all the clinical trials, one can identify all the studies and then address whether selective publication matters. According to the recommendation by the International Committee of Medical Journal Editors (ICMJE) \citep{deangelis2005}, several clinical trial registry systems have been established and widely used in practice such as ClinicalTrials.gov ({\tt https://clinicaltrials.gov/ct2/home}), World Health Organization's (WHO) International Clinical Trials Registry Platform (ICTRP) ({\tt http://apps.who.int/trialsearch/}), EU Clinical Trials Register (EUCTR) ({\tt https//:www.clinicaltrialsregister.euctr-search/search}) and ISRCTN ({\tt https//:www.isrctn.com/}). Actually, the accumulated information in clinical trial registries could potentially be very useful in reducing publication bias \citep{hart2012,baudard2017}. However, their roles in meta-analysis practice are usually limited as a searching tool to identify those conducted but still unpublished studies. Some important study specific information (e.g. the planned sample sizes) in the clinical trial registries has not been utilized efficiently, in particular to address the potential impact on the estimation of effect size. 

\citet{huang2021} utilized the planned sample sizes of studies that were conducted but not published yet, which was available regardless of clinical trial registries, to make inference on the Copas-Shi selection model. \citet{copas2000} proposed to take a sensitivity analysis approach fixing some unknown parameters as sensitivity parameters, since the likelihood function conditional on published was likely to have a flat plateau and was hard to maximize. \citet{huang2021} observed that the full likelihood function with the planned sample size was likely to be convex and all the unknown parameters could be well estimated by maximizing the full likelihood. The method by \citet{huang2021} successfully simplified  the inference for the Copas-Shi selection function. On the other hand, as argued, the Copas-Shi selection function may not be satisfactory in interpretation. In addition, to draw a sound conclusion, it is desirable to evaluate how robust the result is against various settings of the selective publication processes.  

In this paper, we develop a simple inference procedure to correct publication bias under the selective publication process driven by the statistical significance of the result, more specifically, the $t$-type statistic of each study, which is an appealing alternative to the Heckman-type selection function by \citet{copas2000}. We propose a publication bias adjusted estimator based on inverse probability weighting (IPW), which is a widely used technique in missing data problems and causal inference. Considering the correspondence between the propensity score in missing data and causal inference and the selection function in meta-analysis, use of the IPW idea in meta-analysis is very natural and indeed is not new; \citet{matsuoka2007} and \citet{mathur2020} examined the IPW estimator to quantify publication bias in the context of the meta-analysis. However, both relied on sensitivity analysis approaches. That is, the publishing probability which corresponds to the propensity score in the IPW estimator, was pre-defined by the specified selection function and was not calculated from data, which can be a very difficult task in practice. With the planned sample size in the clinical trial registries, we introduce an estimating equation for unknown parameters in the selection function, borrowing the idea to handle the propensity score in the general missing data problem under missing not at random \citep{Kott2010, Miao2016, morikawa2021}. The estimating equation is tractable and once the parameters in the selection function are obtained, our IPW estimator for the overall mean over studies is very simple of a closed form expression. In addition to providing a combined mean, evaluation of the between-study heterogeneity is also an important objective of meta-analyses; the common-effect assumption is implausible in many systematic reviews and therefore random-effects models are recommended in practice \citep{borenstein2010}. We propose an IPW-type DerSimonian-Laird estimator for the between-study variance and also some other heterogeneity measures, all of which have a simple closed form. We developed asymptotic theory and a parametric bootstrap procedure to construct confidence intervals for the overall mean and the between-study variance.

The organization of the rest of the paper is as follows. In Section 2, we introduce notations and the standard DerSimonian-Laird estimator for the random-effect meta-analysis, which our development relied on. In Section 3, the proposed method is introduced. In subsection 3.1, notations considering clinical trial registries are introduced. In Section 3.2, some selection functions based on $t$-type statistics are introduced. In Section 3.3, the IPW estimators for the overall mean and the between-study variance are proposed. In Section 3.4, a parametric bootstrapping for constructing confidence intervals are presented. In Section 3.5, we introduce IPW versions of other heterogeneity measures. In Section 4, we report results of simulation studies to examine the performance of the proposed methods. In Section 5, illustrations are given with some meta-analysis datasets. We conclude this paper by mentioning issues in the methods and potential future work. All the theoretical developments are placed in the web-appendix. 

\section[]{Basic setup and the standard methods for meta-analysis\label{DL}}

Suppose we are conducting a meta-analysis of  \emph{N} published studies to compare two treatment groups. Let the estimated treatment effect of the $i$th study denoted by $y_i$ such as the log-odds ratio or the log-hazard ratio, and its standard error $\sigma_i$ is supposed to be available. Following the standard convention in the meta-analysis field, $\sigma_i$ is assumed to be known in theoretical development. We suppose the following random-effects model; given $\mu_i$ and $\sigma_i$, $y_i\sim N(\mu_i,\sigma_i^2)$. Here, $\mu_i$ is the true value of the $i$th study and is regarded as a random-effect such that $ \mu_i \sim N(\mu,\tau^2)$, where $\mu$ is the treatment effect and  $\tau^2$ is the unknown between-study variance. Then, the marginal model $y_i \sim N(\mu, \sigma_i^2+\tau^2)$ follows from the above. 
 
The inverse variance weighted estimator \citep{cochran1954} for $\mu$ is denoted by
\begin{equation}
\hat{\mu}=\dfrac{\sum_{i=1}^{N}\omega_i y_i}{\sum_{i=1}^{N}\omega_i},
\label{mu}
\end{equation}
where $\omega_i={(\sigma_i^2+\tau^2)}^{-1}$. In practice, $\tau^2$ should be estimated and various estimators are available. In this paper, we consider the DerSimonian-Laird (DL) estimator \citep{dersimonian1986}, which is given by
\begin{equation}
\hat{\tau}_{DL}^2=\max\left\{0,
\dfrac{Q-(N-1)}
{\sum_{i=1}^{N}\sigma_i^{-2}-\sum_{i=1}^{N}\sigma_i^{-4}/\sum_{i=1}^{N}\sigma_i^{-2}
}
\right\},
\label{dl}
\end{equation}
where $Q =\mathlarger{\sum}_{i=1}^{N}(y_i-\hat{\mu}_{F})^2/\sigma_i^2$ is Cochran's Q statistics. $\hat{\mu}_{F}$ is the fixed-effect estimator, which is defined by ($\ref{mu}$) with $\tau^2=0$. 

\section[]{Proposed method\label{IPW}}
\subsection{Clinical trial registry}
In addition to $N$ published studies, suppose we identify $M$ unpublished studies by using clinical trial registries. For $i=1, 2, ..., N+M$, let the random variable $D_i$ be 1 if the $i$th study is published and be 0 if unpublished. Without loss of generality, we assume that the first $N$ studies are published. As defined in Section 2, for published studies, $(y_i, \sigma_i)$ are available. As argued in the introduction, for studies registered in a clinical trial registry, the planned sample sizes of the two groups (not separately by groups) are available regardless of clinical trial registry systems. Let $n_i$ be the number of sample size enrolled in the two groups for published studies and be the planned sample size in the two groups for unpublished studies. We assume $n_i$ is consistent with actual sample size for unpublished studies. Then, we suppose the following data are available; for $i=1, 2,..., N$ (published studies), $(y_i, \sigma_i, n_i)$ is available and for $i=N+1, ..., N+M$ (unpublished studies), only $n_i$ is available. In the following, we suppose $(y_i, \sigma_i, n_i)$ for $i=1, 2, ..., N+M$ are random samples from a population. 

\subsection{Selection functions based on $t$-type statistic}
In this subsection, we introduce some selection functions describing selective publication processes. We focus on the selection functions defined with the $t$-type statistic $t_i=y_i/\sigma_i$. Let the probability to be published of the study with ($y_i, \sigma_i, n_i$) is denoted by $\pi_i(\boldsymbol \beta)=P(D_i=1\mid y_i, \sigma_i, n_i; \boldsymbol  \beta)$, where $\boldsymbol \beta$ is a parameter (vector). We consider one- or two-parameter selection functions. For two-parameter cases, we denote $\boldsymbol \beta=(\beta_0, \beta_1)$.

\citet{preston2004} considered several one-parameter selection functions including the 1-parameter logistic function 
\begin{equation}
 \pi_i(\beta) = \frac{2\exp\left(-\beta\left\{1-\Phi(t_i)\right\}\right)}{1+\exp\left(-\beta\left\{1-\Phi(t_i)\right\}\right)},
\label{logistic1} 
\end{equation}
and the modified 1-parameter logistic function 
\begin{equation}
\pi_i(\beta) = \dfrac{2\exp\left(-\beta \sigma_i\left\{1-\Phi(t_i)\right\}\right)}
{1+\exp\left(-\beta \sigma_i\left\{1-\Phi(t_i)\right\}\right)},
\label{mlogistic1}
\end{equation}
where $\Phi(\cdot)$ is the cumulative function of the standard normal distribution. Other one-parameter selection models were also considered such as the half-normal and the negative-exponential selection functions and their modified versions. \cite{preston2004} proposed to estimate all the parameters of ($\mu, \tau^2, \beta$) by maximizing the conditional log-likelihood function for published studies. However, as they commented, parameters in the selection function might be estimated imprecisely, which in turn may influence the estimates of effect size and result in an unreasonable confidence interval. Probably, due to difficulty in estimation, \citet{preston2004} mainly focused on one-parameter selection functions. Although these one-parameter selection functions have an advantage of simplicity, they have a disadvantage of impossibility to describe the publication process that does not depend on the $t$-type statistic, or say a random selection. If some studies are unpublished independently from outcomes, $\beta$ in the selection function ($\ref{logistic1}$) or (\ref{mlogistic1}) should be zero. Then, the marginal selection probability $p=P(D_i=1)$ should be 1, which does not allow existence of randomly unpublished studies.

Besides, two-parameter selection functions are also considered including the 2-parameter probit model 
\begin{equation}
 \pi_i(\boldsymbol {\beta}) = \Phi(\beta_0+\beta_1 t_i),
\label{probit2}
\end{equation}
and the 2-parameter logistic model
\begin{equation}
 \pi_i(\boldsymbol {\beta}) = 
\dfrac{\exp{(\beta_0+\beta_1 t_i)}}
{1+\exp{(\beta_0+\beta_1 t_i)}}.
\label{logistic2}
\end{equation}

\citet{copas2013} proposed a likelihood-based sensitivity analysis method; with the marginal selection probability $p$ fixed, one could estimate all the parameters by satisfying the marginal selection probability and maximizing the observed conditional likelihood iteratively. Then the impact of the publication bias can be studied by monitoring how the effect size changed as the selection probability decreased.

\subsection{Inverse probability weighting method for publication bias adjustment}
With publication indicator $D_i$, the estimator ($\ref{mu}$) is expressed as
\begin{equation}
\hat{\mu} =\dfrac{\sum_{i=1}^{N}\omega_i y_i}{\sum_{i=1}^{N}\omega_i} 
= \dfrac{\sum_{i=1}^{S}\omega_i D_i y_i}{\sum_{i=1}^{S}\omega_i D_i}
\end{equation}
where $S=N+M$. This representation motivates us to use an estimate of the form
\begin{equation}
\hat{\mu}_{IPW}(\boldsymbol \beta, \tau^2)=\dfrac{\mathlarger{\sum}_{i=1}^{S}
\dfrac{1}{\sigma_i^2+\tau^2}
\dfrac{D_i}{\pi_i(\boldsymbol \beta)} y_i }
{\mathlarger{\sum}_{i=1}^{S}
\dfrac{1}{\sigma_i^2+\tau^2}
\dfrac{D_i}{\pi_i(\boldsymbol \beta)} }.
\label{mu_ipw}
\end{equation}

This is a natural analogy of the inverse probability weighted (IPW) estimator by the propensity score, which is widely used in missing data problems and in causal inference. 
For estimation of $\boldsymbol \beta$, consider the following estimating equation
\begin{equation}
U(\boldsymbol \beta)=\sum_{i=1}^{S}\left\{1-\frac{D_i}{\pi_i(\boldsymbol\beta)}\right\}g(n_i)=0,
\label{eq0}
\end{equation}
where $g(n_i)$ is a function of the same dimension as  $\boldsymbol \beta$. This estimating equation is motivated by the propensity score analysis in the missing not at random setting \citep{Kott2010, Miao2016, morikawa2021}. On specification of $g(n_i)$, one may make an efficiency augment \citep{morikawa2021}, but we employ rather simple ones as follows. When we consider a one-parameter selection function such as ($\ref{logistic1}$) and ($\ref{mlogistic1}$), we use
\begin{equation}
U(\beta)=\sum_{i=1}^{S}\left\{1-\dfrac{D_i}{\pi_i(\beta)}\right\} \sqrt{n_i}=0.
\label{eq1}
\end{equation}

When we use a two-parameter selection function such as ($\ref{probit2}$) and  ($\ref{logistic2}$), we consider the estimating equation,
\begin{equation}
U(\boldsymbol \beta)=\sum_{i=1}^{S}\left\{1-\dfrac{D_i}{\pi_i(\boldsymbol \beta)}\right\} 
\left(
 \begin{array}{c}
      1 \\
      \sqrt{n_i}\\
      \end{array}
\right)=0.
\label{eq2}
\end{equation}

The solution to the equation ($\ref{eq1}$) or ($\ref{eq2}$) is denoted by $\hat{\boldsymbol \beta}$. 
The estimating equations ($\ref{eq1}$) and ($\ref{eq2}$) are unbiased and then $\hat {\boldsymbol \beta}$ consistently estimates the true value $\boldsymbol \beta$ \citep{Kott2010, Miao2016, morikawa2021} if the selection function is correctly specified (see proof in web-appendix A). 

For one-parameter selection functions, one can easily see that ($\ref{eq1}$) is a monotone function of $\beta$ and then the equation can be easily solved by the Newton-Raphson or the binary search methods. 
For two-parameter selection functions, the Hessian matrix for ($\ref{eq2}$) may not be positive definite and we observed computational difficulties in applying the Newton-Raphson method. We propose to obtain the solution to the equation ($\ref{eq2}$) by minimizing 
\begin{equation}
\left|\sum_{i=1}^{S}\left\{1-\dfrac{D_i}{\pi_i(\boldsymbol \beta)}\right\}\right|
+\left|\sum_{i=1}^{S}\left\{1-\dfrac{D_i}{\pi_i(\boldsymbol \beta)}\right\}\sqrt{n_i}\right|.
\end{equation}
We use the nlminb() function in \emph{R} (package stats, version 3.6.2) for implementation.   

For estimation of $\tau^2$, we propose an IPW version of 
the DL estimator, which is defined by $\hat{\tau}_{IPW}^2=\hat{\tau}_{IPW}^2(\hat{\boldsymbol \beta})$, where
\begin{equation}
\hat{\tau}_{IPW}^2(\boldsymbol \beta)=\max\left\{0,\dfrac{Q_{IPW}(\boldsymbol \beta)-\{S-1\}}
     {\mathlarger{\sum}_{i=1}^{S}\dfrac{1}{\sigma_i^2}\dfrac{D_i}{\pi_i(\boldsymbol \beta)}-A_S(\boldsymbol \beta)/B_S(\boldsymbol \beta)}\right \},
\label{tau_ipw}
\end{equation}
$A_S(\boldsymbol \beta)=S^{-1}\mathlarger{\sum}_{i=1}^{S}\dfrac{1}{\sigma_i^4}\dfrac{D_i}{\pi_i(\boldsymbol \beta)}$,
$B_S(\boldsymbol \beta)=S^{-1}\mathlarger{\sum}_{i=1}^{S}\dfrac{1}{\sigma_i^2}\dfrac{D_i}{\pi_i(\boldsymbol \beta)}$,
\begin{eqnarray*}
Q_{IPW}(\boldsymbol \beta) =\sum_{i=1}^{S}\dfrac{1}{\sigma_i^2}\dfrac{D_i}{\pi_i(\boldsymbol \beta)}
\left\{y_i-\hat{\mu}_{F, IPW}(\boldsymbol \beta) \right\}^2,
\label{q_ipw} 
\end{eqnarray*}
and 
\begin{eqnarray*}
\hat{\mu}_{F, IPW}(\boldsymbol \beta) &=\dfrac{\mathlarger{\sum}_{i=1}^{S}\dfrac{1}{\sigma_i^2}\dfrac{D_i}{\pi_i(\boldsymbol \beta)} y_i}
{\mathlarger{\sum}_{i=1}^{S}\dfrac{1}{\sigma_i^2}\dfrac{D_i}{\pi_i(\boldsymbol \beta)}}. 
\label{fix_ipw} 
\end{eqnarray*}
 We call the estimator ($\ref{tau_ipw}$) the IPW-DL estimator. $Q_{IPW}(\boldsymbol \beta)$ and $\hat{\mu}_{F, IPW}(\boldsymbol \beta)$ are the IPW versions of \emph{Q} statistics in ($\ref{dl}$) and the fixed-effect model estimator, respectively.  

Finally, we propose the IPW estimator $\hat{\mu}_{IPW}=\hat{\mu}_{IPW}(\hat{\boldsymbol\beta}, \hat{\tau}_{IPW}^2)$ for $\mu$. In web-appendix A, we show consistency of $\hat{\mu}_{IPW}$ and $\hat{\tau}_{IPW}^2$ if the selection function is correctly specified as $S$ goes to infinity and $n_i$ goes to infinity for each $i$. Confidence intervals of $\mu$, $\tau^2$ as well as $\boldsymbol\beta$, can be constructed with the consistent estimators of their asymptotic variance, whose derivations and definitions are given in web-appendix B. 

\subsection{Parametric bootstrap confidence intervals}
Alternatively, one may use a parametric bootstrap approach to construct confidence intervals. Conditional on the data, parametric bootstrap samples $\tilde{y}_i$ are generated from $\tilde{y}_i \sim N(\hat{\mu}_{IPW}, \sigma_i^2+\hat{\tau}_{IPW}^2)$ \citep{turner2000, viechtbauer2007}. 
Define
\begin{equation*}
\tilde{U}(\boldsymbol \beta)=\sum_{i=1}^{S}\left\{1-\frac{D_i}{\tilde{\pi}_i(\boldsymbol\beta)}\right\}g(n_i)=0.
\end{equation*}
where $\tilde{\pi}_i(\boldsymbol \beta)$ is defined by $\pi_i(\boldsymbol \beta)$ replacing $t_i=y_i/\sigma_i$ with $\tilde{y}_i/\sigma_i$,
Let the solution to $\tilde{U}(\boldsymbol \beta)=0$ denoted by 
$\tilde{\boldsymbol \beta}$.  
Define $\tilde{\tau}_{IPW}^2=\tilde{\tau}_{IPW}^2(\tilde{\boldsymbol \beta})$, where
\begin{equation*}
\tilde{\tau}_{IPW}^2(\tilde{\boldsymbol \beta})=\max\left\{0,\dfrac{\tilde{Q}_{IPW}(\tilde{\boldsymbol \beta})-\{S-1\}}
     {\mathlarger{\sum}_{i=1}^{S}\dfrac{1}{\sigma_i^2}\dfrac{D_i}{\tilde{\pi}_i(\tilde{\boldsymbol \beta})}-\tilde{A}_S(\tilde{\boldsymbol \beta})/\tilde{B}_S(\tilde{\boldsymbol \beta})} \right\}
\label{tau_IPW_tilde}
\end{equation*}
$\tilde{A}_S(\tilde{\boldsymbol \beta})=S^{-1}\mathlarger{\sum}_{i=1}^{S}\dfrac{1}{\sigma_i^4}\dfrac{D_i}{\tilde{\pi}_i(\tilde{\boldsymbol \beta})}$,
$\tilde{B}_S(\tilde{\boldsymbol \beta})=S^{-1}\mathlarger{\sum}_{i=1}^{S}\dfrac{1}{\sigma_i^2}\dfrac{D_i}{\tilde{\pi}_i(\tilde{\boldsymbol \beta})}$,
\begin{eqnarray*}
\tilde{\mu}_{F, IPW}(\tilde{\boldsymbol \beta}) &=\dfrac{\mathlarger{\sum}_{i=1}^{S}\dfrac{1}{\sigma_i^2}\dfrac{D_i}{\tilde{\pi}_i(\tilde{\boldsymbol \beta})} \tilde{y}_i}
{\mathlarger{\sum}_{i=1}^{S}\dfrac{1}{\sigma_i^2}\dfrac{D_i}{\tilde{\pi}_i(\tilde{\boldsymbol \beta})}}, 
\end{eqnarray*}

and
\begin{eqnarray*}
\tilde{Q}_{IPW}(\tilde{\boldsymbol \beta}) &=\mathlarger{\sum}_{i=1}^{S}\dfrac{1}{\sigma_i^2}\dfrac{D_i}{\tilde{\pi}_i(\tilde{\boldsymbol \beta})}\left\{\tilde{y}_i-\tilde{\mu}_{F, IPW}(\tilde{\boldsymbol \beta})\right\}^2, 
\end{eqnarray*}

Then, define $\tilde{\mu}_{IPW}=\tilde{\mu}_{IPW}(\tilde{\boldsymbol \beta},\tilde{\tau}^2_{IPW})$, where
\begin{eqnarray*}
\tilde{\mu}_{IPW}(\tilde{\boldsymbol \beta}, \tilde{\tau}^2_{IPW})&=\dfrac{\mathlarger{\sum}_{i=1}^{S}\dfrac{D_i}{\tilde{\pi}_i(\tilde{\boldsymbol \beta})}\dfrac{1}{\sigma_i^2+\tilde{\tau}^2_{IPW}} \tilde{y}_i}{\mathlarger{\sum}_{i=1}^{S}\dfrac{D_i}{\tilde{\pi}_i(\tilde{\boldsymbol \beta})}\dfrac{1}{\sigma_i^2+\tilde{\tau}^2_{IPW}}},
\end{eqnarray*}

For $i=1, 2, ..., S$, sufficiently large number (say, 1000) of parametric bootstrap samples of $\tilde{y}_i$ are generated. Let the number of bootstrap samples denoted by $B$ and the $b$th bootstrap sample is denoted by $\tilde{y}_i^{(b)}$. Denote $\tilde{\mu}_{IPW}$ with the $b$th bootstrap samle by $\tilde{\mu}_{IPW}^{(b)}$. Define the bootstrap variance for $\mu$ by $\sigma_{boot}^2=B^{-1} \sum_{b=1}^B (\tilde{\mu}_{IPW}^{(b)}-\bar{\mu}_{boot})$, where $\bar{\mu}_{boot}=B^{-1} \sum_{b=1}^B \tilde{\mu}_{IPW}^{(b)}$ and a bootstrap two-tailed 95 percent confidence interval is constructed by $\hat{\mu}_{IPW}+q(0.025) \sigma_{boot}, \hat{\mu}_{IPW}+q(0.975) \sigma_{boot}$, where $q(0.025)$ and $q(0.975)$ are the 2.5 and 97.5 percentiles of the standardized bootstrap samples of $ (\tilde{\mu}_{IPW}^{(b)}-\bar{\mu}_{boot})/\sigma_{boot}$. Bootstrap confidence intervals of $\tau^2$ based on $\hat{\tau}_{IPW}^2$ are constructed in a similar way.

\subsection{Other measures of between-study heterogeneity}
\citet{higgins2002} discussed several heterogeneity measures alternative to $\tau^2$, including $H^2=Q/N-1$ and $I^2=(H^2-1)/H^2$. The former can be interpreted approximately as the ratio of confidence interval widths for the overall mean from random-effects and fixed-effect models, the latter can be used to describe the percentage of  variability for $\mu$ that is due to heterogeneity rather than sampling error. The $I^2$ has been adopted by the Cochrane Collaboration as the summary measure of heterogeneity in their Review Manager Software and other commonly used packages for meta-analysis (e.g. metafor package, meta package). With the IPW version of $Q$-statistics ($Q_{IPW}$), the IPW versions of $H^2$ and $I^2$ can be defined as $H_{IPW}^2=Q_{IPW}/(S-1)$ and $I_{IPW}^2=(H^2_{IPW}-1)/H^2_{IPW}$, which would be useful to describe heterogeneity in the presence of selective publication process.

\section{Simulation study\label{sim}}
\subsection{Settings}
Simulation studies were carried out to assess the performance of the proposed IPW estimator. We conducted two kinds of simulation studies; one was based on one-parameter selection functions and the other on two-parameter ones. We generate multiple studies and according to one- or two-parameter selection functions, some of them were selected as published studies. 

We begin with describing how to generate complete data of published and unpublished studies. The simulation design for generating all the studies was similar to those considered in \citet{huang2021}. Suppose we are interested in conducting a meta-analysis of randomized clinical trials to compare two treatment groups with a dichotomous outcome. The log-odds ratio was used as the summary measure of the treatment effect between the experimental group and control group. We set the population treatment effect $\mu$ = -0.50 which was motivated by the Clopidogrel study in Section 5.2 and $\tau$ = 0.05, 0.15 or 0.30, which reflects small to moderate heterogeneity. The total number of studies including published and unpublished was set as 15, 25, 50 or 100. At first, we generated the true log-odds ratio of the $i$th study $\mu_i$ from $N(\mu,\tau^2)$. Next, we generated the true event rate in the control group \emph{$p_{ic}$} from the uniform distribution \emph{U (0.2,0.9)} and then the event rate in the treatment group \emph{$p_{it}$} can be derived as $e^{\mu_i}p_{ic}/(1-p_{ic}+p_{ic}e^{\mu_i})$. Following \citet{kuss2015}, the total sample size of each study was generated from \emph{LN(5,1)}, the log-normal distribution with the location parameter 5 and scale parameter 1, and the minimum sample size was restricted to 20 patients (values below 20 were rounded up to 20). Subjects were allocated to the two treatment groups with probability of 0.5. Then the individual participant data could be generated from the binomial distributions \emph{B($n_{ic},p_{ic}$)} and \emph{B($n_{it},p_{it}$)}, respectively. With the generated individual participant data, we could calculate the empirical log odds ratio $y_i$ and its standard error $\sigma_i$. 

\begin{figure}
\centerline{%
\includegraphics[width=10cm, height=8.5cm]{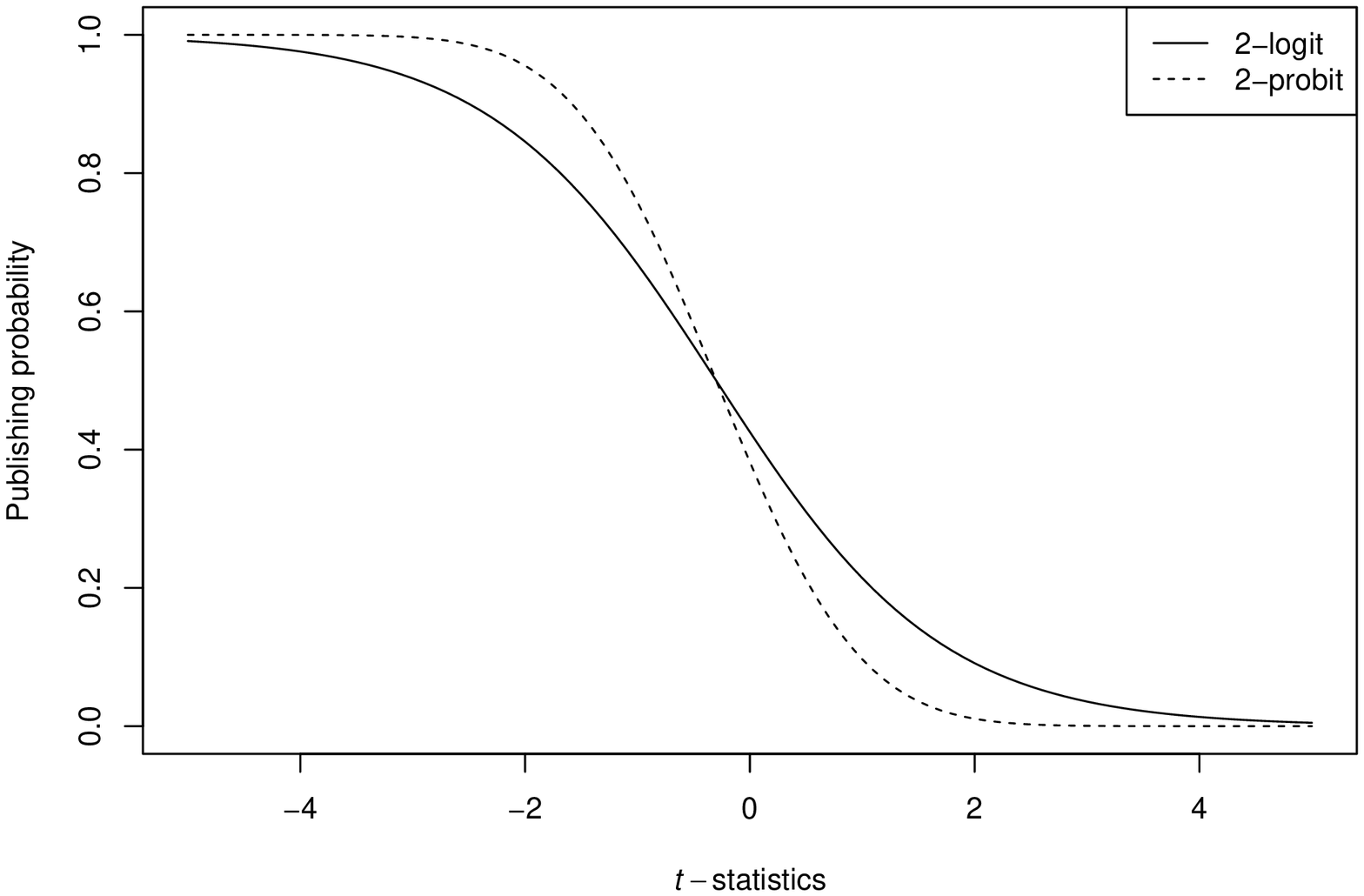}}
\caption{Plot of the two-parameter selection models used to generate simulation datasets}
\label{fig1}
\end{figure}

From the complete data generated following the above procedure, we selectively picked several studies according to one- or two-parameter selection models and then created four datasets, which are referred as {\it sDatasets 1} to {\it 4}, among which the first two were based on one-parameter selection functions and the latter two were on two-parameter ones. The indicator of publication status \emph{$D_i$} was generated from the binomial distribution \emph{$B(1,\pi_i(\boldsymbol\beta))$}. For {\it sDataset 1}, we selected published studies with the one-parameter logistic selection function ($\ref{logistic1}$) of $\beta=2$. For {\it sDataset 2}, the one-parameter modified logistic selection function ($\ref{mlogistic1}$) of $\beta=5$ was used. In these datasets, about 20 percent studies were regarded as unpublished. For {\it sDataset 3} and {\it sDataset 4}, the two-parameter selection functions of ($\ref{probit2}$) and ($\ref{logistic2}$) with $\boldsymbol \beta$=(-0.3, -1) were used, and about 25 percent studies in {\it sDataset 3} and 30 percent studies in {\it sDataset 4} were regarded as unpublished, respectively. Selection functions used to generate {\it sDataset 3} and {\it sDataset 4} were plotted in Figure~\ref{fig1}.

\subsection{Results with one-parameter selection functions}
In this subsection, we summarize results for one-parameter selection functions. In estimation, we used the one-parameter logistic selection function ($\ref{logistic1}$) and the modified logistic selection function ($\ref{mlogistic1}$). For {\it sDataset 1}, the logistic selection model was correctly specified and the modified one was mis-specified. For {\it sDataset 2} vise versa. We examined influence of correct/mis-specification of the selection function on estimation. For comparison, we applied the maximum conditional likelihood method by \citet{preston2004} with a correctly-specified or mis-specified selection function. To maximize the conditional log-likelihood, we used the nlminb() function in \emph{R}. 

In Table~\ref{1par.mu}, we presented the simulation results for estimation of $\mu$ for {\it sDataset 1}. The results for {\it sDataset 2} were presented in the web-supplementary Table S1. We applied the standard mixed-effects model ($\ref{mu}$) with the DerSimonian-Laird $\tau^2$ estimator using metafor package in \emph{R} and observed that it had considerable biases. We found that the maximum conditional likelihood method by \citet{preston2004} failed to converge in about 20 percent realizations. Furthermore, even if the selection function was correctly specified, there were still certain biases and the coverage probabilities were far from the nominal level of 95 percent. 

On the other hand, the proposed IPW estimator successfully obtained estimates in all the realizations. If the selection function was correctly specified, the IPW estimator eliminated publication biases and the proposed asymptotic confidence intervals had empirical coverage probabilities close to the nominal level of 95 percent under the large study scenarios (S = 50 and 100), while the parametric bootstrap confidence intervals can result in much improvement with few studies (S = 15 and 25). For {\it sDataset 1}, misspecification of the selection function did not lead serious biases. For {\it sDataset 2}, as summarized in the web-supplementary Table S1, we observed that misspecification led certain biases with large number of studies (S = 50 and 100). 

Results for estimation of $\tau^2$ were presented in Table~\ref{1par.tau} and the web-supplementary Table S2 for {\it sDataset 1} and  {\it sDataset 2}, respectively. 
We observed that the DerSimonian-Laird estimator $\tau_{DL}^2$ may substantially underestimate the heterogeneity due to the selective publication process and the proportion of zero $\tau^2$ estimates could be extremely high even when $S$= 50 and 100, similar findings were also reported by \citet{augusteijn2019} and \citet{friede2017}; while our IPW version of the DerSimonian-Laird estimator $\tau_{IPW}^2$ had smaller biases and less zero estimates in most scenarios. For both {\it sDataset 1} and {\it sDataset 2}, misspecification of the selection function did not influence the performance so much. However, the coverage probabilities of the asymptotic confidence intervals for the $\tau_{IPW}^2$  estimator were not necessarily close to the nominal level for large $\tau^2$, whereas the parametric bootstrap confidence intervals led more conservative coverage probabilities.

\subsection{Results with two-parameter selection functions}
In this subsection, we summarized the results with the two-parameter selection functions. For {\it sDataset 3}, the two-parameter probit model was correctly specified and the two-parameter logistic model was misspecified, and for {\it sDataset 4} vise versa. We compared our proposed method with the maximum conditional likelihood method by \citet{copas2013}. As mentioned in Section 3.2, the method is implemented with a marginal selection probability fixed (sensitivity analysis). In order to make a fair comparison, we used the empirical publication rate ($p=N/S$) in implementation of the Copas method, and $nlminb()$ function was used for its conditional log-likelihood optimization.

In Table~\ref{2par.mu}, we presented the simulation results of $\mu$ estimates with {\it sDataset 3}, and the results for {\it sDataset 4} were presented in the web-supplementary Table S3. For reference, we also showed results with the standard mixed-effects model. The crude estimates were highly biased suggesting that the simulation design successfully generated data under selective publication. Both the Copas sensitivity analysis method and the proposed IPW method could reduce the biases and ours had smaller biases in almost all the  scenarios when the selection model was correctly specified. We observed that the profile likelihood method in the Copas sensitivity analysis gave substantially narrow confidence intervals of inaccurate coverage probabilities. The asymptotic confidence intervals for the IPW estimator might be so wide. On the other hand, the confidence intervals based on parametric bootstrap seemed more reasonable and the coverage probabilities were close to the nominal level in almost all the scenarios. We also observed that both in {\it sDataset 3} and {\it sDataset 4}, mis-specification of selection function could introduce considerable biases, although the mis-specified IPW estimators were still less biased  than the standard mixed-effect model.

We presented the simulation results of $\tau^2$ estimates for {\it sDataset 3} and {\it sDataset 4} in Tables~\ref{2par.tau} and web-supplementary S4, respectively. We observed that our IPW version of DerSimonian-Laird $\tau_{IPW}^2$ estimator had smaller bias and less zero estimates than the $\tau_{DL}^2$ estimator in most scenarios. Although the coverage probabilities of the asymptotic confidence intervals were unsatisfactory when the true $\tau$ was 0.3, a more conservative parametric bootstrap confidence interval can always perform well with the coverage probabilities close to the nominal level of 95 percent. We also observed that mis-specification of the selection function did not have much impact on the performance of $\tau_{IPW}^2$  in both {\it sDataset 3} and {\it sDataset 4}.

\section{Examples\label{exam}}
\subsection{Antidepressant study}
Firstly, we illustrate our proposed method with the antidepressant study which aimed to evaluate the improvement in depression symptoms of 12 antidepressant drugs, and the outcome was measured as the standardized mean difference between the treatment group and placebo group. In this study, \citet{turner2008} identified 73 registered randomized clinical trials from the FDA registry, among them 50 were published and 23 were unpublished, and selective publication process was suggested by the nature of data that most of the published studies showed statistical significance while unpublished studies did not (see \citet{turner2008} for more details). Since their focus was the meta-analysis of studies used for licensing, only the FDA registry was used for study searching and hence both the effect size and standard error were available for all the studies (published and unpublished). Although this was not a typical situation of meta-analysis, we used this dataset for an illustrative purpose of our proposed method. Regarding the overall mean of all the 73 studies with the standard mixed-effect model as the "gold standard", we compared the performance of our proposed method and other competitive methods empirically. The “gold standard” of DerSimonian-Laird estimate with all the 73 studies was 0.344 with a 95\% CI of [0.300, 0.388], while the DerSimonian-Laird estimate only with the 50 published studies was 0.409 with a 95\% CI of [0.366, 0.453], indicating that the underlying selective publication process might have considerable influence on estimation (see Table~\ref{turner}). 

At first, we summarized the results with the one-parameter selection functions.  We applied the one-parameter logistic ($\ref{logistic1}$)  and its modified version ($\ref{mlogistic1}$), the $\hat\beta$ were estimated as 7.168 (95\% asymptotic CI: [3.106, 11.231]; 95\% bootstrap CI: [6.158, 8.711]) and 47.722 (95\% asymptotic CI: [21.524, 73.920]; 95\% bootstrap CI: [42.743, 55.285]) with ($\ref{logistic1}$) and ($\ref{mlogistic1}$), respectively. The resulting estimates of $\mu$ as well as those conditional likelihood-based estimators were summarized in Table~\ref{turner}.  Preston's conditional likelihood-based method gave the estimates of 0.355 (95\% CI: [0.296, 0.414]) and 0.357 (95\% CI: [0.301, 0.414]) with the one-parameter logistic selection function ($\ref{logistic1}$) and its modified version ($\ref{mlogistic1}$), respectively. Our IPW method gave the more conservative estimates as 0.333 (95\% asymptotic CI: [0.283, 0.383]; 95\% bootstrap CI: [0.263, 0.395]) and 0.339 (95\% CI: [0.287, 0.392]; 95\% bootstrap CI: [0.251, 0.411]), accordingly.

\begin{figure}
\centerline{%
\includegraphics[width=16cm, height=8.5cm]{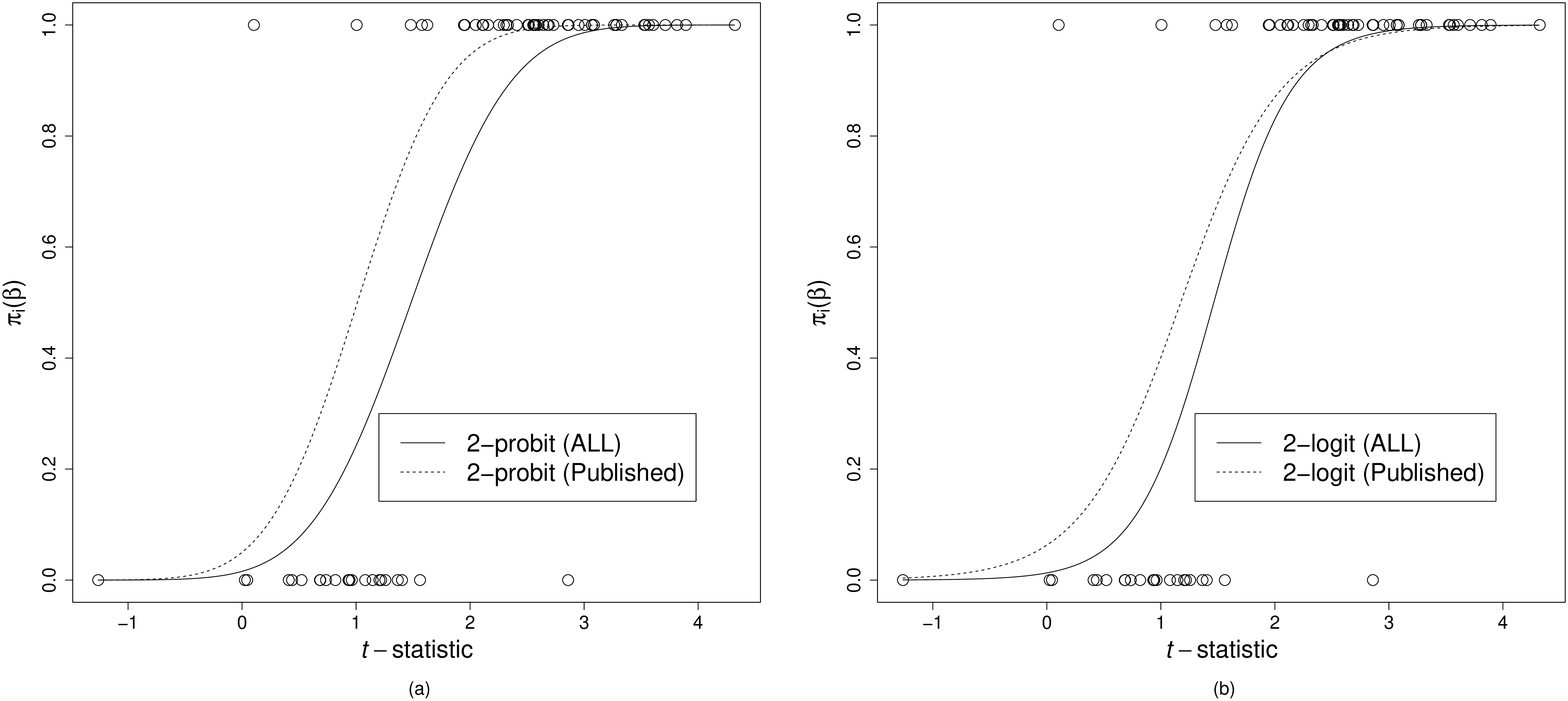}}
\caption{Plot of estimated selective publication processes for the Antidepressant study: Two-parameter probit model ($\Phi(-1.645+1.627t_i)$); Two-parameter logistic model ($\frac{\exp(-2.706+2.290t_i)}{1+\exp(-2.706+2.290t_i)}$)  }
\label{fig2}
\end{figure}

Next, we demonstrated the results with the two-parameter probit ($\ref{probit2}$) and logistic ($\ref{logistic2}$) selection functions. As we mentioned in last paragraph, one benefit of this data is it included all the information for both published and unpublished studies, hence an empirical comparison could be done by checking the estimation of $\boldsymbol{\hat\beta}$=$(\hat\beta_0, \hat\beta_1)$ using standard maximum likelihood estimation (MLE) applied to all the 73 studies and our estimating equations ($\ref{eq2}$) to the 50 published studies. For two-parameter probit ($\ref{probit2}$) selection function, the estimated selection functions were plotted with solid line and dashed line in Figure~\ref{fig2} (a) for MLE and our method, respectively. For the estimation using MLE, $\hat\beta_0=-2.151$ (95\% CI: [-3.206, -1.223]) and $\hat\beta_1=1.488$ (95\% CI: [0.979, 2.097]); as to our estimation simply using the sample sizes of unpublished studies, we got $\hat\beta_0=-1.645$ (95\% asymptotic CI: [-18.158, 14.867]; 95\% bootstrap CI: [-2.379, -1.117]) and $\hat\beta_1=1.627$ (95\% asymptotic CI: [-9.122, 12.375]; 95\% bootstrap CI: [0.995, 2.046]). The asymptotic CIs were very wide, while the bootstrap ones seemed relevant. Observations for the two-parameter logistic ($\ref{logistic2}$) selection function were similar to this (Figure~\ref{fig2} (b)). We explained such observations in simulation studies, and we trust the bootstrap CIs more. With both selection functions, the null hypothesis of $\beta_1=0$ was statistically significant, successfully suggesting a selective publication process behind. For the results of $\mu$ estimates with two-parameter selection functions, we estimated the Copas selection model with the marginal selection probability fixed at $p=50/73$ and obtained the estimate of 0.373 with a very short 95\% CI of [0.356, 0.405]. Our IPW method gave the estimates of 0.330 (95\% asymptotic CI: [0.282, 0.378]; 95\% bootstrap CI: [0.219, 0.419]) and 0.339 (95\% asymptotic CI: [0.295, 0.383]; 95\% bootstrap CI: [0.258, 0.400]) with the two-parameter probit ($\ref{probit2}$) and logistic ($\ref{logistic2}$) selection function, respectively. It seemed that in this study all these methods successfully eliminate certain publication bias.

We also compared the estimation of heterogeneity with the methods above. We observed that all the methods only relying on published studies gave zero estimates, while the proposed IPW version of DerSimonian-Laird $\hat{\tau}_{IPW}^2$ estimator gave the non-zero estimates. With 73 studies (published and unpublished), the $I^2$ was 22.8\%. On the other hand, with only published 50 studies, it was estimated as 0\%, whereas the IPW version of $I^2$ ranged from 34.8\% to 38.4\% with different selection functions (see Table~\ref{turner}). 

\subsection{Clopidogrel study}
\citet{chen2013} conducted a meta-analysis of 12 published studies to compare the high and standard maintenance-dose clopidogrel on major adverse cardiovascular/cerebrovascular events (MACE/MACCE). \citet{huang2021} revisited this study and identified 3 unpublished studies from multiple clinical trial registries (see Table S5 in web-appendix D for details). We use this data to gain some insights of the performance of our IPW method in small meta-analysis. 

We first illustrated the proposed method using one-parameter logistic selection function ($\ref{logistic1}$) and its modified version ($\ref{mlogistic1}$), $\hat\beta$ were estimated as 1.018 (95\% asymptotic CI: [-0.222, 2.257]; 95\% bootstrap CI: [0.611, 1.681]) and 1.309 (95\% asymptotic CI: [-0.114, 2.733]; 95\% bootstrap CI: [0.953, 1.957]), respectively. The results of $\mu$ estimates were presented in Table~\ref{clopidogrel}. Without accounting for the publication bias, the result of standard mixed-effects model concluded the significantly lower event rate in the high maintenance-dose clopidogrel group with the pooled odds ratio of 0.622 and a 95\% CI of [0.441, 0.877]. While the adjusted results with these one-parameter selection functions suggested that the significant effect of high maintenance-dose of clopidogrel might be marginal. Furthermore, the estimates with Preston's conditional likelihood method were very sensitive to the choice of the selection functions which was similar to observations in the simulation study; the integrated odds ratios were estimated as 0.849 (95\% CI: [0.319, 2.259]) and 0.696 (95\% CI: [0.434, 1.116]) with the one-parameter logistic ($\ref{logistic1}$) and its modified version ($\ref{mlogistic1}$), respectively. In contrast, the IPW estimates with these two selection functions were relatively close; the pooled odds ratio were estimated as 0.666 (95\% asymptotic CI: [0.452, 0.982]; 95\% bootstrap CI: [0.471, 0.953]) and 0.648 (95\% asymptotic CI: [0.425, 0.987]; 95\% bootstrap CI: [0.451, 0.965]), respectively.

\begin{figure}[b]
\centerline{%
\includegraphics[width=10cm, height=8.5cm]{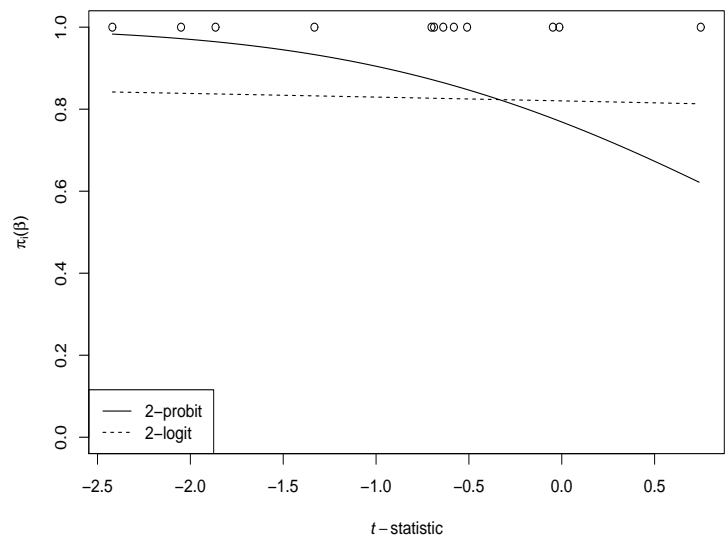}}
\caption{Plot of estimated selective publication processes for the Clopidogrel study: Two-parameter probit model ($\Phi(0.735-0.575t_i)$); Two-parameter logistic model ($\frac{\exp(1.518-0.064t_i)}{1+\exp(1.518-0.064t_i)}$)    }
\label{fig3}
\end{figure}

Next, we demonstrated the results with the two-parameter selection functions. The estimated selection functions were shown in Figure~\ref{fig3}. We observed an almost flat dotted line with $\hat\beta_1$ estimated as -0.064 for the two-parameter logistic ($\ref{logistic2}$) selection function, indicating that it might be failed to identify the selective publication process; while the two-parameter probit ($\ref{probit2}$) selection function gave the estimate of $\hat\beta_1$ as -0.575 (95\% asymptotic CI: [-4.104, 2.954]; 95\% bootstrap CI: [-1.119, 0.153]), although we still could not reject the null hypothesis of $\beta_1=0$, in Figure~\ref{fig3} the solid line indicated that the selective publication process might be concerned. Similar with the antidepressant study, we found the bootstrap CI might be more reasonable for the $\boldsymbol\beta$ inference in practice. For the estimation of $\mu$, Copas sensitivity analysis method gave the pooled odds ratio as 0.691 and a 95\% CI of [0.468, 1.012] with the marginal selection probability fixed at $p=12/15$; while the proposed IPW method with two-parameter probit ($\ref{probit2}$) gave the estimate of 0.662 with a 95\% asymptotic CI of [0.474, 0.923] and a 95\% bootstrap CI of [0.468, 0.904]. As we observed in Figure~\ref{fig3}, two-parameter logistic ($\ref{logistic2}$) selection function did not suggest the selective publication process. Then the resulting estimate was very close to the standard mixed-effects model (see Table~\ref{clopidogrel}). 

In summary, we must be cautious of the failure in estimating the selection function for small meta-analysis, and then plotting the selection functions and checking the estimate of $\boldsymbol{\hat\beta}$ will be helpful in practice. On the other hand, all the $\hat{\tau}_{IPW}^2$ were 0, while the conditional likelihood-based methods gave a moderate heterogeneity. Similarly, \citet{huang2021} also reported that the methods using maximum likelihood estimation with the 12 published studies gave a moderate heterogeneity, while the publication bias adjustment method with all the 15 studies gave a zero estimate.

\section{Discussion\label{dis}}

In this paper, we successfully introduced the IPW method to address the publication bias issue in meta-analysis context. Differently from  \citet{matsuoka2007} and \citet{mathur2020}, by introducing a simple estimating equation for the selection function, we can avoid massy processes of sensitivity analyses. The simplicity and flexibility of the IPW estimator allows us to handle various $t$-type selection functions, and as shown in Section~\ref{sim}, it can result in certain improvement in estimating both overall effect size and heterogeneity than the original conditional likelihood-based methods by \citet{preston2004} and \citet{copas2013}. On the other hand, we focus on one- and two-parameter selection functions in this paper, since the information of unpublished studies from clinical trial registries only enables us to handle small number of parameters. Selection functions with more parameters \citep{dear1992, hedges1992} might be useful to describe more flexible and complicated selective publication processes. It would be worthwhile to develop methods to handle such kind of selection functions.

Publication bias issue has long been recognized as a kind of missing data problem. However, there is a notable difference between the publication bias issue and the general missing data problem. In general missing data problems such as drop-out in clinical trials, the whole study population is clearly understood. In other words, we know how many subjects are missing and some information such as baseline covariates are available for missing subjects. In the publication bias issue, it is hard to define a complete study population since we only observed published studies. Due to this reason, well-developed missing data methodologies such as the IPW method are hard to be used in this area directly and most of the methods for publication bias rely on funnel-plot symmetry. After long years development of clinical trial registries, prospective registration has been widely accepted by clinical trial researchers, and searching on clinical trial registries plays a more and more important role when performing systematic reviews. This allows us to identify those unpublished studies and give us the opportunity to handle the publication bias issue like a general missing data problem.

In our view, clinical trial registries play an important role to fill the gap between the publication bias issue and the general missing data problem. Our development of the IPW estimator as well as the maximum likelihood estimation by \citet{huang2021} was along with this perspective. These two methods used different types of selection functions and then complement each other. With these methods, we can address robustness of the results of meta-analysis against different selective publication process described by the Heckman-type and the $t$-type selection functions. Since it was impossible to identify the true selective publication process in reality, a comprehensive sensitivity analysis with multiple selection functions would be useful and is always recommended in practice.

\clearpage
\newpage

\newpage

\newpage

\clearpage

\begin{sidewaystable}
\begin{center}
\caption{Simulation results for estimation of $\mu$ under one-parameter logistic selection model with $\beta=2$ and $\tau$ = 0.05, 0.15 or 0.30  \label{1par.mu}}
\scalebox{0.6}[1]{%
\begin{tabular}{*{4}{l}*{16}{c}}
\hline
&&&&\multicolumn{4}{c}{$S=15$}&\multicolumn{4}{c}{$S=25$}&\multicolumn{4}{c}{$S=50$}&\multicolumn{4}{c}{$S=100$}\\
\cmidrule(lr){5-8}\cmidrule(lr){9-12}\cmidrule(lr){13-16}\cmidrule(lr){17-20}
$\tau^2$&Method&Selection&Status&AVE(SD)&CP&LOCI&NOC&AVE(SD)&CP&LOCI&NOC&AVE(SD)&CP&LOCI&NOC&AVE(SD)&CP&LOCI&NOC\\
\hline
0.0025      
 & DL &  &  & -0.546 ( 0.081 ) & 0.940 & 0.341 & 1000 & -0.535 ( 0.061 ) & 0.934 & 0.253 & 1000 & -0.531 ( 0.044 ) & 0.896 & 0.173 & 1000 & -0.531 ( 0.031 ) & 0.826 & 0.120 & 1000\\

 & Preston  & 1-logit & C & -0.370 ( 0.600 ) & 0.759 & 8.378 & 816 & -0.416 ( 0.336 ) & 0.764 & 3.180 & 828 & -0.469 ( 0.144 ) & 0.785 & 0.170 & 833 & -0.486 ( 0.041 ) & 0.784 & 0.121 & 834\\

 & & 1-mlogit & M & -0.484 ( 0.132 ) & 0.829 & 0.322 & 877 & -0.496 ( 0.072 ) & 0.831 & 0.236 & 869 & -0.502 ( 0.050 ) & 0.824 & 0.165 & 847 & -0.506 ( 0.035 ) & 0.811 & 0.111 & 827\\

 & IPW (Asym) & 1-logit & C & -0.509 ( 0.087 ) & 0.872 & 0.281 & 1000 & -0.503 ( 0.065 ) & 0.915 & 0.226 & 1000 & -0.497 ( 0.046 ) & 0.923 & 0.166 & 1000 & -0.499 ( 0.033 ) & 0.921 & 0.119 & 1000\\

 & IPW(Boot) & 1-logit & C & -0.509 ( 0.087 ) & 0.971 & 0.369 & 1000 & -0.503 ( 0.065 ) & 0.959 & 0.270 & 1000 & -0.497 ( 0.046 ) & 0.960 & 0.184 & 1000 & -0.499 ( 0.033 ) & 0.939 & 0.127 & 1000\\

 & IPW (Asym) & 1-mlogit & M & -0.512 ( 0.090 ) & 0.877 & 0.285 & 1000 & -0.506 ( 0.066 ) & 0.913 & 0.229 & 1000 & -0.500 ( 0.048 ) & 0.920 & 0.169 & 1000 & -0.502 ( 0.035 ) & 0.921 & 0.120 & 1000\\

 & IPW(Boot) & 1-mlogit & M & -0.512 ( 0.090 ) & 0.976 & 0.392 & 1000 & -0.506 ( 0.066 ) & 0.963 & 0.288 & 1000 & -0.500 ( 0.048 ) & 0.966 & 0.200 & 1000 & -0.502 ( 0.035 ) & 0.953 & 0.139 & 1000\\

\hline
0.0225     
 & DL &  &  & -0.551 ( 0.097 ) & 0.897 & 0.363 & 1000 & -0.552 ( 0.076 ) & 0.872 & 0.281 & 1000 & -0.544 ( 0.052 ) & 0.842 & 0.191 & 1000 & -0.542 ( 0.036 ) & 0.760 & 0.133 & 1000\\

 &Preston   & 1-logit & C & -0.350 ( 0.536 ) & 0.689 & 27.627 & 804 & -0.409 ( 0.346 ) & 0.683 & 11.959 & 796 & -0.449 ( 0.195 ) & 0.718 & 12.713 & 793 & -0.477 ( 0.073 ) & 0.745 & 6.243 & 809\\

 &  & 1-mlogit & M & -0.482 ( 0.143 ) & 0.782 & 0.337 & 834 & -0.501 ( 0.116 ) & 0.747 & 1.269 & 819 & -0.508 ( 0.059 ) & 0.805 & 4.070 & 847 & -0.511 ( 0.042 ) & 0.810 & 8.056 & 830\\

 & IPW (Asym) & 1-logit & C & -0.507 ( 0.102 ) & 0.864 & 0.323 & 1000 & -0.504 ( 0.080 ) & 0.884 & 0.270 & 1000 & -0.498 ( 0.056 ) & 0.923 & 0.198 & 1000 & -0.496 ( 0.037 ) & 0.934 & 0.143 & 1000\\

& IPW(Boot) & 1-logit & C & -0.507 ( 0.102 ) & 0.938 & 0.391 & 1000 & -0.504 ( 0.080 ) & 0.931 & 0.299 & 1000 & -0.498 ( 0.056 ) & 0.927 & 0.203 & 1000 & -0.496 ( 0.037 ) & 0.931 & 0.142 & 1000\\

 & IPW (Asym) & 1-mlogit & M & -0.511 ( 0.106 ) & 0.863 & 0.328 & 1000 & -0.508 ( 0.081 ) & 0.888 & 0.275 & 1000 & -0.501 ( 0.060 ) & 0.920 & 0.201 & 1000 & -0.500 ( 0.040 ) & 0.932 & 0.146 & 1000\\

& IPW(Boot) & 1-mlogit & M & -0.511 ( 0.106 ) & 0.942 & 0.414 & 1000 & -0.508 ( 0.081 ) & 0.952 & 0.323 & 1000 & -0.501 ( 0.060 ) & 0.941 & 0.221 & 1000 & -0.500 ( 0.040 ) & 0.947 & 0.155 & 1000\\

\hline
0.0900     
 & DL &  &  & -0.592 ( 0.121 ) & 0.844 & 0.454 & 1000 & -0.590 ( 0.092 ) & 0.807 & 0.350 & 1000 & -0.588 ( 0.064 ) & 0.722 & 0.250 & 1000 & -0.586 ( 0.046 ) & 0.530 & 0.178 & 1000\\

 &Preston   & 1-logit & C & -0.327 ( 0.579 ) & 0.649 & 33.151 & 793 & -0.397 ( 0.361 ) & 0.669 & 57.391 & 767 & -0.409 ( 0.266 ) & 0.653 & 12.507 & 759 & -0.439 ( 0.157 ) & 0.647 & 24.018 & 751\\

 &  & 1-mlogit & M & -0.496 ( 0.192 ) & 0.767 & 48.400 & 801 & -0.514 ( 0.147 ) & 0.737 & 5.634 & 829 & -0.528 ( 0.089 ) & 0.760 & 17.064 & 841 & -0.534 ( 0.058 ) & 0.754 & 9.569 & 846\\

 & IPW (Asym) & 1-logit & C & -0.511 ( 0.137 ) & 0.864 & 0.429 & 1000 & -0.505 ( 0.101 ) & 0.918 & 0.356 & 1000 & -0.501 ( 0.071 ) & 0.923 & 0.262 & 1000 & -0.497 ( 0.050 ) & 0.943 & 0.192 & 1000\\

& IPW(Boot) & 1-logit & C & -0.511 ( 0.137 ) & 0.916 & 0.486 & 1000 & -0.505 ( 0.101 ) & 0.935 & 0.370 & 1000 & -0.501 ( 0.071 ) & 0.918 & 0.258 & 1000 & -0.497 ( 0.050 ) & 0.939 & 0.182 & 1000\\

 & IPW (Asym) & 1-mlogit & M & -0.515 ( 0.142 ) & 0.859 & 0.441 & 1000 & -0.508 ( 0.109 ) & 0.906 & 0.365 & 1000 & -0.503 ( 0.080 ) & 0.910 & 0.272 & 1000 & -0.500 ( 0.057 ) & 0.928 & 0.198 & 1000\\

& IPW(Boot) & 1-mlogit & M & -0.515 ( 0.142 ) & 0.922 & 0.523 & 1000 & -0.508 ( 0.109 ) & 0.946 & 0.407 & 1000 & -0.503 ( 0.080 ) & 0.930 & 0.287 & 1000 & -0.500 ( 0.057 ) & 0.933 & 0.203 & 1000\\

\hline
                                         &&&True&-0.500&-&-&-&-0.500&-&-&-&-0.500&-&-&-&-0.500&-&-&-\\
\hline
\end{tabular}}
\flushleft {Selection, the selection model used for estimation: 1-logit denotes the one-parameter logistic selection model, 1-mlogit denotes the one-parameter modified logistic selection model; Status, model specification: C means selection model correctly specified, M means selection model misspecified; s, number of total studies; AVE, mean value of estimates; SD, standard error of estimates; CP, 95\%confidence interval coverage probability; LOCI, length of confidence interval; NOC, number of converged cases; DL, random-effects model with DerSimonian-Laird method; Preston, Preston's conditional likelihood method; IPW (Asym), the proposed method with asymptotic variance; IPW (Boot), the proposed method with parametric bootstrap confidence interval}
\end{center}
\end{sidewaystable}

\begin{sidewaystable}
\begin{center}
\caption{Simulation results for estimation of $\tau^2$ under one-parameter logistic selection model with $\beta=2$ and $\tau$ = 0.05, 0.15 or 0.30\label{1par.tau}}
\scalebox{0.6}[1]{%
\begin{tabular}{*{4}{l}*{16}{c}}
\hline
&&&&\multicolumn{4}{c}{$S=15$}&\multicolumn{4}{c}{$S=25$}&\multicolumn{4}{c}{$S=50$}&\multicolumn{4}{c}{$S=100$}\\
\cmidrule(lr){5-8}\cmidrule(lr){9-12}\cmidrule(lr){13-16}\cmidrule(lr){17-20}
$\tau^2$&Method&Selection&Status&AVE(SD)&CP&LOCI&NOZ&AVE(SD)&CP&LOCI&NOZ&AVE(SD)&CP&LOCI&NOZ&AVE(SD)&CP&LOCI&NOZ\\
\hline
0.0025       

& DL &  &  & 0.009 ( 0.020 ) & 0.952 & 0.235 & 669 & 0.006 ( 0.013 ) & 0.936 & 0.117 & 663 & 0.004 ( 0.008 ) & 0.908 & 0.054 & 687 & 0.002 ( 0.004 ) & 0.895 & 0.026 & 749\\

 & IPW (Asym) & 1-logit & C & 0.012 ( 0.031 ) & 0.992 & 0.067 & 653 & 0.010 ( 0.023 ) & 0.997 & 0.054 & 607 & 0.008 ( 0.015 ) & 0.990 & 0.042 & 549 & 0.006 ( 0.009 ) & 0.995 & 0.031 & 535\\

& IPW(Boot) & 1-logit & C & 0.012 ( 0.031 ) & 0.998 & 0.131 & 653 & 0.010 ( 0.023 ) & 0.997 & 0.089 & 607 & 0.008 ( 0.015 ) & 0.997 & 0.059 & 549 & 0.006 ( 0.009 ) & 0.995 & 0.039 & 535\\

 & IPW (Asym) & 1-mlogit & M & 0.013 ( 0.036 ) & 0.993 & 0.069 & 645 & 0.011 ( 0.033 ) & 0.995 & 0.057 & 601 & 0.010 ( 0.022 ) & 0.987 & 0.046 & 530 & 0.009 ( 0.015 ) & 0.986 & 0.037 & 500\\

& IPW(Boot) & 1-mlogit & M & 0.013 ( 0.036 ) & 0.998 & 0.146 & 645 & 0.011 ( 0.033 ) & 0.996 & 0.105 & 601 & 0.010 ( 0.022 ) & 0.990 & 0.076 & 530 & 0.009 ( 0.015 ) & 0.976 & 0.055 & 500\\

\hline
0.0225   
 & DL &  &  & 0.018 ( 0.029 ) & 0.944 & 0.287 & 492 & 0.017 ( 0.024 ) & 0.942 & 0.168 & 427 & 0.013 ( 0.016 ) & 0.918 & 0.085 & 332 & 0.011 ( 0.011 ) & 0.870 & 0.050 & 256\\

 & IPW (Asym) & 1-logit & C & 0.023 ( 0.039 ) & 0.967 & 0.085 & 488 & 0.024 ( 0.031 ) & 0.963 & 0.078 & 369 & 0.022 ( 0.023 ) & 0.946 & 0.063 & 238 & 0.021 ( 0.018 ) & 0.927 & 0.052 & 138\\

& IPW(Boot) & 1-logit & C & 0.023 ( 0.039 ) & 0.999 & 0.155 & 488 & 0.024 ( 0.031 ) & 1.000 & 0.121 & 369 & 0.022 ( 0.023 ) & 0.996 & 0.083 & 238 & 0.021 ( 0.018 ) & 0.992 & 0.061 & 138\\

 & IPW (Asym) & 1-mlogit & M & 0.024 ( 0.042 ) & 0.968 & 0.087 & 497 & 0.025 ( 0.033 ) & 0.963 & 0.081 & 380 & 0.024 ( 0.030 ) & 0.941 & 0.068 & 252 & 0.024 ( 0.023 ) & 0.923 & 0.058 & 145\\

& IPW(Boot) & 1-mlogit & M & 0.024 ( 0.042 ) & 0.999 & 0.169 & 497 & 0.025 ( 0.033 ) & 1.000 & 0.138 & 380 & 0.024 ( 0.030 ) & 0.991 & 0.101 & 252 & 0.024 ( 0.023 ) & 0.979 & 0.078 & 145\\

\hline
0.0900   
  & DL &  &  & 0.059 ( 0.063 ) & 0.947 & 0.489 & 214 & 0.058 ( 0.050 ) & 0.929 & 0.278 & 113 & 0.058 ( 0.035 ) & 0.871 & 0.165 & 33 & 0.058 ( 0.026 ) & 0.778 & 0.105 & 4\\

 & IPW (Asym) & 1-logit & C & 0.072 ( 0.078 ) & 0.645 & 0.166 & 211 & 0.077 ( 0.066 ) & 0.734 & 0.158 & 85 & 0.080 ( 0.047 ) & 0.791 & 0.137 & 19 & 0.083 ( 0.035 ) & 0.827 & 0.113 & 2\\

& IPW(Boot) & 1-logit & C & 0.072 ( 0.078 ) & 0.929 & 0.281 & 211 & 0.077 ( 0.066 ) & 0.927 & 0.234 & 85 & 0.080 ( 0.047 ) & 0.913 & 0.180 & 19 & 0.083 ( 0.035 ) & 0.921 & 0.136 & 2\\

 & IPW (Asym) & 1-mlogit & M & 0.073 ( 0.082 ) & 0.658 & 0.171 & 220 & 0.078 ( 0.069 ) & 0.733 & 0.163 & 98 & 0.081 ( 0.052 ) & 0.781 & 0.143 & 25 & 0.083 ( 0.038 ) & 0.834 & 0.117 & 1\\

& IPW(Boot) & 1-mlogit & M & 0.073 ( 0.082 ) & 0.945 & 0.299 & 220 & 0.078 ( 0.069 ) & 0.952 & 0.250 & 98 & 0.081 ( 0.052 ) & 0.949 & 0.193 & 25 & 0.083 ( 0.038 ) & 0.951 & 0.148 & 1\\
\hline
\end{tabular}}
\flushleft {Selection, the selection model used for estimation: 1-logit denotes the one-parameter logistic selection model, 1-mlogit denotes the one-parameter modified logistic selection model; Status, model specification: C means selection model correctly specified, M means selection model misspecified; s, number of total studies; AVE, mean value of estimates; SD, standard error of estimates; CP, 95\%confidence interval coverage probability; LOCI, length of confidence interval; NOZ, number of 0 estimates; DL, random-effects model with DerSimonian-Laird method;  IPW (Asym), the proposed method with asymptotic variance; IPW (Boot), the proposed method with parametric bootstrap confidence interval}
\end{center}
\end{sidewaystable}

\begin{sidewaystable}
\begin{center}
\caption{Simulation results for estimation of $\mu$ under two-parameter probit selection model with $\boldsymbol\beta=(-0.3,-1.0)$ and $\tau$ = 0.05, 0.15 or 0.30\label{2par.mu}}
\scalebox{0.6}[1]{%
\begin{tabular}{*{4}{l}*{16}{c}}
\hline
&&&&\multicolumn{4}{c}{$S=15$}&\multicolumn{4}{c}{$S=25$}&\multicolumn{4}{c}{$S=50$}&\multicolumn{4}{c}{$S=100$}\\
\cmidrule(lr){5-8}\cmidrule(lr){9-12}\cmidrule(lr){13-16}\cmidrule(lr){17-20}
$\tau^2$&Method&Selection&Status&AVE(SD)&CP&LOCI&NOZ&AVE(SD)&CP&LOCI&NOZ&AVE(SD)&CP&LOCI&NOZ&AVE(SD)&CP&LOCI&NOZ\\
\hline
0.0025 
& DL &  &  & -0.552 ( 0.083 ) & 0.944 & 0.346 & 1000 & -0.547 ( 0.064 ) & 0.910 & 0.256 & 1000 & -0.545 ( 0.044 ) & 0.858 & 0.176 & 1000 & -0.543 ( 0.030 ) & 0.726 & 0.121 & 1000\\

 & Copas & 2-probit & C & -0.511 ( 0.091 ) & 0.539 & 0.187 & 958 & -0.503 ( 0.073 ) & 0.563 & 0.154 & 990 & -0.498 ( 0.049 ) & 0.634 & 0.117 & 996 & -0.498 ( 0.034 ) & 0.673 & 0.085 & 999\\

 & IPW (Asym) & 2-logit & M & -0.524 ( 0.081 ) & 0.907 & 0.840 & 1000 & -0.519 ( 0.063 ) & 0.920 & 1.678 & 1000 & -0.518 ( 0.043 ) & 0.943 & 0.473 & 1000 & -0.519 ( 0.030 ) & 0.928 & 0.183 & 1000\\

 & IPW(Boot) & 2-logit & M & -0.524 ( 0.081 ) & 0.971 & 0.361 & 1000 & -0.519 ( 0.063 ) & 0.961 & 0.269 & 1000 & -0.518 ( 0.043 ) & 0.954 & 0.192 & 1000 & -0.519 ( 0.030 ) & 0.949 & 0.139 & 1000\\

 & IPW (Asym) & 2-probit & C &  -0.508 ( 0.086 ) & 0.902 & 0.772 & 1000 & -0.502 ( 0.068 ) & 0.914 & 1.586 & 1000 & -0.497 ( 0.049 ) & 0.953 & 1.215 & 1000 & -0.497 ( 0.034 ) & 0.971 & 0.676 & 1000\\

& IPW(Boot) & 2-probit & C & -0.508 ( 0.086 ) & 0.974 & 0.391 & 1000 & -0.502 ( 0.068 ) & 0.966 & 0.296 & 1000 & -0.497 ( 0.049 ) & 0.972 & 0.221 & 1000 & -0.497 ( 0.034 ) & 0.987 & 0.176 & 1000\\

\hline
0.0225      
& DL &  &  & -0.569 ( 0.098 ) & 0.879 & 0.369 & 1000 & -0.565 ( 0.078 ) & 0.837 & 0.280 & 1000 & -0.561 ( 0.052 ) & 0.734 & 0.190 & 1000 & -0.560 ( 0.036 ) & 0.582 & 0.132 & 1000\\

 & Copas & 2-probit & C & -0.522 ( 0.113 ) & 0.475 & 0.184 & 947 & -0.509 ( 0.093 ) & 0.526 & 0.158 & 990 & -0.506 ( 0.063 ) & 0.539 & 0.115 & 998 & -0.503 ( 0.043 ) & 0.605 & 0.087 & 999\\

 & IPW (Asym) & 2-logit & M & -0.531 ( 0.096 ) & 0.903 & 0.975 & 1000 & -0.526 ( 0.077 ) & 0.912 & 28.010 & 1000 & -0.524 ( 0.053 ) & 0.933 & 0.918 & 1000 & -0.525 ( 0.036 ) & 0.934 & 0.331 & 1000\\

 & IPW(Boot) & 2-logit & M & -0.531 ( 0.096 ) & 0.940 & 0.381 & 1000 & -0.526 ( 0.077 ) & 0.921 & 0.294 & 1000 & -0.524 ( 0.053 ) & 0.927 & 0.208 & 1000 & -0.525 ( 0.036 ) & 0.918 & 0.152 & 1000\\

 & IPW (Asym) & 2-probit & C & -0.514 ( 0.100 ) & 0.895 & 0.906 & 1000 & -0.504 ( 0.082 ) & 0.924 & 1.225 & 1000 & -0.500 ( 0.059 ) & 0.954 & 0.683 & 1000 & -0.498 ( 0.042 ) & 0.970 & 0.816 & 1000\\

 & IPW(Boot) & 2-probit & C & -0.514 ( 0.100 ) & 0.952 & 0.417 & 1000 & -0.504 ( 0.082 ) & 0.946 & 0.334 & 1000 & -0.500 ( 0.059 ) & 0.954 & 0.246 & 1000 & -0.498 ( 0.042 ) & 0.972 & 0.194 & 1000\\

\hline
0.0900      
 & DL &  &  & -0.627 ( 0.126 ) & 0.790 & 0.452 & 1000 & -0.621 ( 0.096 ) & 0.712 & 0.341 & 1000 & -0.623 ( 0.067 ) & 0.510 & 0.246 & 1000 & -0.620 ( 0.046 ) & 0.231 & 0.176 & 1000\\

 & Copas & 2-probit & C & -0.564 ( 0.154 ) & 0.373 & 0.197 & 973 & -0.551 ( 0.121 ) & 0.396 & 0.167 & 983 & -0.534 ( 0.094 ) & 0.392 & 0.118 & 999 & -0.518 ( 0.067 ) & 0.440 & 0.093 & 1000\\

 & IPW (Asym) & 2-logit & M & -0.56 ( 0.126 ) & 0.878 & 1.638 & 1000 & -0.552 ( 0.093 ) & 0.927 & 1.203 & 1000 & -0.551 ( 0.067 ) & 0.937 & 1.603 & 1000 & -0.549 ( 0.047 ) & 0.940 & 1.094 & 1000\\

& IPW(Boot) & 2-logit & M & -0.560 ( 0.126 ) & 0.893 & 0.462 & 1000 & -0.552 ( 0.093 ) & 0.908 & 0.354 & 1000 & -0.551 ( 0.067 ) & 0.879 & 0.261 & 1000 & -0.549 ( 0.047 ) & 0.851 & 0.198 & 1000\\

 & IPW (Asym) & 2-probit & C & -0.538 ( 0.131 ) & 0.872 & 1.677 & 1000 & -0.523 ( 0.101 ) & 0.937 & 1.267 & 1000 & -0.514 ( 0.075 ) & 0.965 & 2.740 & 1000 & -0.507 ( 0.055 ) & 0.982 & 1.931 & 1000\\

& IPW(Boot) & 2-probit & C & -0.538 ( 0.131 ) & 0.913 & 0.506 & 1000 & -0.523 ( 0.101 ) & 0.951 & 0.403 & 1000 & -0.514 ( 0.075 ) & 0.956 & 0.309 & 1000 & -0.507 ( 0.055 ) & 0.978 & 0.250 & 1000\\

\hline
                                         &&&True&-0.500&-&-&-&-0.500&-&-&-&-0.500&-&-&-&-0.500&-&-&-\\
\hline
\end{tabular}}
\flushleft {Selection, the selection model used for estimation: 2-logit denotes the two-parameter logistic selection model, 2-probit denotes the two-parameter probit selection model; Status, model specification: C means selection model was correctly specified, M means selection model was misspecified; s, number of total studies; AVE, mean value of estimates; SD, standard error of estimates; CP, 95\%confidence interval coverage probability; LOCI, length of confidence interval; NOC, number of converged cases; DL, random-effects model with DerSimonian-Laird method; Copas, Copas' sensitivity analysis method; IPW (Asym), the proposed method with asymptotic variance; IPW (Boot), the proposed method with parametric bootstrap confidence interval}
\end{center}
\end{sidewaystable}

\begin{sidewaystable}
\begin{center}
\caption{Simulation results for estimation of $\tau^2$ under two-parameter probit selection model with $\boldsymbol\beta=(-0.3,-1.0)$ and $\tau$ = 0.05, 0.15 or 0.30\label{2par.tau}}
\scalebox{0.6}[1]{%
\begin{tabular}{*{4}{l}*{16}{c}}
\hline
&&&&\multicolumn{4}{c}{$S=15$}&\multicolumn{4}{c}{$S=25$}&\multicolumn{4}{c}{$S=50$}&\multicolumn{4}{c}{$S=100$}\\
\cmidrule(lr){5-8}\cmidrule(lr){9-12}\cmidrule(lr){13-16}\cmidrule(lr){17-20}
$\tau^2$&Method&Selection&Status&AVE(SD)&CP&LOCI&NOZ&AVE(SD)&CP&LOCI&NOZ&AVE(SD)&CP&LOCI&NOZ&AVE(SD)&CP&LOCI&NOZ\\
\hline
0.0025 
& DL &  &  & 0.007 ( 0.018 ) & 0.937 & 0.240 & 714 & 0.005 ( 0.011 ) & 0.925 & 0.108 & 714 & 0.003 ( 0.006 ) & 0.910 & 0.048 & 734 & 0.001 ( 0.003 ) & 0.842 & 0.021 & 826\\

 & IPW (Asym) & 2-logit & M & 0.007 ( 0.018 ) & 0.999 & 0.099 & 740 & 0.005 ( 0.011 ) & 0.999 & 0.227 & 732 & 0.003 ( 0.009 ) & 0.999 & 0.070 & 771 & 0.001 ( 0.004 ) & 1.000 & 0.032 & 826\\

 & IPW(Boot) & 2-logit & M & 0.007 ( 0.018 ) & 1.000 & 0.113 & 740 & 0.005 ( 0.011 ) & 1.000 & 0.080 & 732 & 0.003 ( 0.009 ) & 1.000 & 0.058 & 771 & 0.001 ( 0.004 ) & 0.999 & 0.044 & 824\\

 & IPW (Asym) & 2-probit & C & 0.009 ( 0.023 ) & 0.996 & 0.111 & 710 & 0.008 ( 0.017 ) & 0.997 & 0.268 & 660 & 0.007 ( 0.017 ) & 0.995 & 0.219 & 620 & 0.005 ( 0.011 ) & 0.998 & 0.145 & 596\\

 & IPW(Boot) & 2-probit & C & 0.009 ( 0.023 ) & 0.998 & 0.135 & 710 & 0.008 ( 0.017 ) & 1.000 & 0.101 & 660 & 0.007 ( 0.017 ) & 0.990 & 0.081 & 621 & 0.005 ( 0.011 ) & 0.994 & 0.071 & 595\\

\hline
0.0225 
  & DL &  &  & 0.015 ( 0.027 ) & 0.934 & 0.287 & 552 & 0.013 ( 0.020 ) & 0.927 & 0.143 & 514 & 0.009 ( 0.014 ) & 0.866 & 0.071 & 450 & 0.008 ( 0.010 ) & 0.784 & 0.040 & 384\\

 & IPW (Asym) & 2-logit & M & 0.014 ( 0.026 ) & 0.962 & 0.126 & 600 & 0.012 ( 0.021 ) & 0.958 & 2.100 & 538 & 0.010 ( 0.015 ) & 0.917 & 0.120 & 506 & 0.008 ( 0.012 ) & 0.823 & 0.060 & 424\\

& IPW(Boot) & 2-logit & M & 0.014 ( 0.026 ) & 1.000 & 0.132 & 600 & 0.012 ( 0.021 ) & 1.000 & 0.102 & 538 & 0.010 ( 0.015 ) & 1.000 & 0.072 & 506 & 0.008 ( 0.011 ) & 1.000 & 0.056 & 423\\

 & IPW (Asym) & 2-probit & C & 0.017 ( 0.031 ) & 0.961 & 0.132 & 571 & 0.017 ( 0.027 ) & 0.963 & 0.206 & 476 & 0.016 ( 0.022 ) & 0.959 & 0.133 & 383 & 0.018 ( 0.020 ) & 0.930 & 0.172 & 202\\

& IPW(Boot) & 2-probit & C & 0.017 ( 0.031 ) & 1.000 & 0.157 & 571 & 0.017 ( 0.027 ) & 1.000 & 0.133 & 475 & 0.016 ( 0.022 ) & 0.996 & 0.103 & 383 & 0.018 ( 0.020 ) & 0.982 & 0.089 & 202\\

\hline
0.0900 
  & DL &  &  & 0.047 ( 0.057 ) & 0.918 & 0.458 & 294 & 0.044 ( 0.042 ) & 0.894 & 0.241 & 187 & 0.045 ( 0.032 ) & 0.802 & 0.143 & 67 & 0.046 ( 0.023 ) & 0.600 & 0.091 & 7\\

 & IPW (Asym) & 2-logit & M &0.045 ( 0.056 ) & 0.600 & 0.244 & 329 & 0.046 ( 0.044 ) & 0.639 & 0.208 & 188 & 0.049 ( 0.036 ) & 0.650 & 0.324 & 72 & 0.051 ( 0.027 ) & 0.647 & 0.222 & 12\\

 & IPW(Boot) & 2-logit & M & 0.045 ( 0.056 ) & 0.884 & 0.223 & 329 & 0.046 ( 0.044 ) & 0.859 & 0.179 & 188 & 0.049 ( 0.036 ) & 0.842 & 0.144 & 72 & 0.051 ( 0.027 ) & 0.829 & 0.118 & 12\\

 & IPW (Asym) & 2-probit & C & 0.049 ( 0.060 ) & 0.625 & 0.349 & 321 & 0.054 ( 0.050 ) & 0.707 & 0.284 & 162 & 0.062 ( 0.044 ) & 0.777 & 0.572 & 50 & 0.068 ( 0.035 ) & 0.826 & 0.441 & 4\\

& IPW(Boot) & 2-probit & C & 0.049 ( 0.060 ) & 0.917 & 0.253 & 321 & 0.054 ( 0.050 ) & 0.917 & 0.214 & 162 & 0.062 ( 0.044 ) & 0.956 & 0.179 & 50 & 0.068 ( 0.035 ) & 0.961 & 0.148 & 4\\
\hline
\end{tabular}}
\flushleft {Selection, the selection model used for estimation: 2-logit denotes the two-parameter logistic selection model, 2-probit denotes the two-parameter probit selection model; Status, model specification: C means selection model correctly specified, M means selection model misspecified; s, number of total studies; AVE, mean value of estimates; SD, standard error of estimates; CP, 95\%confidence interval coverage probability; LOCI, length of confidence interval; NOZ, number of 0 estimates; DL, random-effects model with DerSimonian-Laird method;  IPW (Asym), the proposed method with asymptotic variance; IPW (Boot), the proposed method with parametric bootstrap confidence interval}    
\end{center}
\end{sidewaystable}

\begin{sidewaystable}
\centering
\caption{Summary of the statistical analysis for publication bias evaluation of Antidepressant study}
\label{turner}
\scalebox{0.75}[1]{%
\begin{tabular}{llllccccc}
 \hline
 Description & Data & Method  &Selection& $\mu$ (95\% CI) &$P$-value  & $\tau^2$ (95\% CI) &$I^2$\\
 \hline
No adjustment &Published \& Unpublished&  DL  &-&0.344 [0.300, 0.388]&$<$0.001& 0.008 [0.000, 0.027]&0.228\\ 
                    &Published &  DL  &-&0.409 [0.366, 0.453]&$<$0.001& 0.000 [0.000, 0.009]&0.000\\
\hline
One-parameter 
 &Published&Preston&1-logit&0.355 [0.296, 0.414]&$<$0.001&0.000 [0.000, 0.016]&-\\
 &Published&Preston&1-mlogit&0.357 [0.301, 0.414]&$<$0.001&0.000 [0.000, 0.016]&-\\
&Published \& Registry&IPW (Asym)&1-logit&0.333 [0.283, 0.383]&$<$0.001&0.017 [0.006, 0.027]&0.376\\
&Published \& Registry&IPW (Boot)&1-logit&0.333 [0.264, 0.395]&-&0.017 [0.000, 0.050]&0.376\\
 &Published \& Registry&IPW (Asym)& 1-mlogit&0.339 [0.287, 0.392]&$<$0.001&0.015 [0.003, 0.027]&0.348\\ 
&Published \& Registry&IPW (Boot)&1-mlogit&0.339 [0.251, 0.411]&-&0.015 [0.000, 0.060]&0.348\\

\hline
Two-parameter 
&Published&Copas&2-probit&0.373 [0.356, 0.405]&-&\multicolumn{1}{l}{0.000} &-\\
&Published \& Registry&IPW(Asym)  &2-probit&0.330 [0.282, 0.378]&$<$0.001&0.017 [0.006, 0.028]&0.384\\
&Published \& Registry&IPW(Boot)  &2-probit&0.330 [0.219, 0.419]&-&0.017 [0.000, 0.069]&0.384\\
&Published \& Registry&IPW(Asym)  &2-logit&0.339 [0.295, 0.383]&$<$0.001&0.015 [0.004, 0.026]&0.353\\
&Published \& Registry&IPW(Boot)  &2-logit&0.339 [0.258, 0.400]&-&0.015 [0.000, 0.050]&0.353\\
\hline
\end{tabular}}
\flushleft {Preston, Preston's conditional likelihood method; Copas, Copas' sensitivity analysis method; IPW (Asym), the proposed IPW method using asymptotic variance; IPW (Boot), the proposed IPW method using parametric bootstrap confidence interval; 1-logit, the one-parameter logistic selection model, 1-mlogit, the one-parameter modified logistic selection model; 2-probit, the two-parameter probit selection model; 2-logit, the two-parameter logistic selection model}
\end{sidewaystable}

\begin{sidewaystable}
\centering
\caption{Summary of the statistical analysis for publication bias evaluation of Clopidogrel study}
\label{clopidogrel}
\scalebox{0.75}[1]{%
\begin{tabular}{lllcccc}
 \hline
 Description & Method  & Selection &OR (95\% CI) &$P$-value& $\tau^2$ (95\% CI) &$I^2$ \\
 \hline
No adjustment &  DL  &-&0.622 [0.441, 0.877]&0.007&0.000 [0.000, 0.754]&0.000\\ 
\hline
One-parameter 
&Preston&1-logit&0.849 [0.319, 2.259]&0.732&0.076 [0.000, 0.461]&-\\
 &Preston&1-mlogit&0.696 [0.434, 1.116]&0.052 &0.045 [0.000, 0.287]&-\\
&IPW (Asym)&1-logit&0.666 [0.452, 0.982]&0.040 &0.000 [0.000, 0.181]&0.000\\
&IPW (Boot)&1-logit&0.666 [0.471, 0.953]&- &0.000 [0.000, 0.463]&0.000\\
 &IPW (Asym)&1-mlogit&0.648 [0.425, 0.987]&0.044&0.000 [0.000, 0.202]&0.000\\
 &IPW (Boot)&1-mlogit&0.648 [0.451, 0.965]&-&0.000 [0.000, 0.534]&0.000\\
\hline
Two-parameter 
&Copas&2-probit&0.691 [0.468, 1.012]&-&\multicolumn{1}{l}{0.092}&-\\
 &IPW (Asym)&2-probit&0.662 [0.474, 0.923]&0.015&0.000 [0.000, 0.183]&0.000\\
 &IPW (Boot)&2-probit&0.662 [0.468, 0.904]&-&0.000 [0.000, 0.354]&0.000\\
 &IPW (Asym)&2-logit&0.625 [0.416, 0.939]&0.024&0.000 [0.000, 0.222]&0.000\\
 &IPW (Boot)&2-logit&0.625 [0.457, 0.861]&-&0.000 [0.000, 0.342]&0.000\\
\hline
\end{tabular}}
\flushleft {Preston, Preston's conditional likelihood method; Copas, Copas' sensitivity analysis method; IPW (Asym), the proposed IPW method using asymptotic variance; IPW (Boot), the proposed IPW method using parametric bootstrap confidence interval; 1-logit, the one-parameter logistic selection model, 1-mlogit, the one-parameter modified logistic selection model; 2-probit, the two-parameter probit selection model; 2-logit, the two-parameter logistic selection model}
\end{sidewaystable}
\clearpage
\newpage

\appendix
\begin{center}
\Large 
Web-appendix to ``Adjusting for publication bias in meta-analysis via inverse probability weighting using clinical trial registries''
\end{center}

\section*{Appendix A: Consistency of $\hat{\tau}_{IPW}^2$ and $\hat{\mu}_{IPW}$}
Suppose the selection function $\pi_i(\boldsymbol \beta)$ is correctly specified and the true value of $\boldsymbol\beta$ is denoted by $\boldsymbol\beta_*$. We assume that $E\Big\{1-D_i/\pi_i(\boldsymbol\beta)\Big\}g(n_i)=\boldsymbol{0}$ has a unique solution.

By the uniform law of large number, it holds that
$\frac{1}{S}U^{\boldsymbol\beta}(\boldsymbol\beta)\xrightarrow{P}E\left[\Big\{1-D_i/\pi_i(\boldsymbol\beta)\Big\}g(n_i)\right]$
uniformly in $\boldsymbol\beta$. By simple algebra,

\begin{equation}
E\left[\Big\{1-\frac{D_i}{\pi_i(\boldsymbol\beta_*)}\Big\}g(n_i)\right]=E\left[g(n_i)\right]-E\left[\frac{g(n_i)}{\pi_i (\boldsymbol\beta_*)}E(D_i\mid y_i,\sigma_i,n_i)\right]=\boldsymbol{0}.
\label{eq_con}
\end{equation}
Then, from the assumption of the uniqueness of the solution to ($\ref{eq_con}$), by theorem 5.9 of van der Vaart \cite{van2000}, one can show the consistency of $\hat{\beta}$ to $\beta_*$.  

Next we show $\hat{\tau}_{IPW}^2 \overset{p}{\to} \tau^2$. Since $A_S(\boldsymbol{\hat\beta})$ and $B_S(\boldsymbol{\hat\beta})$ converge in probability to some constants, it holds that
\begin{align*}
\hat{\tau}_{IPW}^2
&=\frac{\frac{1}{S}Q_{IPW}(\boldsymbol{\hat\beta})-1+S^{-1} }
{\frac{1}{S}\sum_{i=1}^{S}\frac{1}{\sigma_i^2}\frac{D_i}{\pi_i(\hat{\boldsymbol\beta})}-S^{-1} A_S(\boldsymbol{\hat\beta})/B_S(\boldsymbol{\hat\beta})
}
\simeq \frac{\frac{1}{S}\sum_{i=1}^S\frac{1}{\sigma_i^2}\frac{D_i}{\pi_i (\boldsymbol \beta_*)}\left\{y_i-\hat\mu_{F, IPW}(\boldsymbol \beta_*)\right\}^2-1}{\frac{1}{S}\sum_{i=1}^{S}\frac{1}{\sigma_i^2}\frac{D_i}{\pi_i(\boldsymbol \beta_*)}}
\nonumber \\
& \simeq \frac{\frac{1}{S}\sum_{i=1}^{S}\frac{1}{\sigma_i^2}\frac{D_i}{\pi_i(\boldsymbol \beta_*)}(y_i-\mu)^2-1}{\frac{1}{S}\sum_{i=1}^{S}\frac{1}{\sigma_i^2}\frac{D_i}{\pi_i (\boldsymbol \beta_*)}}
\overset{P}{\to} \frac{E\left[\frac{1}{\sigma_i^2}(\sigma_i^2+\tau^2)\right]-1}{E\left[\frac{1}{\sigma_i^2}\right]}=\tau^2
\end{align*}

Similarly, it holds that
\begin{align*}
\hat{\mu}_{IPW}&=\frac{\sum_{i=1}^{S} \frac{D_i}{\pi_i(\hat{\boldsymbol\beta})}\frac{1}{\sigma_i^2+\hat{\tau}_{IPW}^2} y_i}{\sum_{i=1}^{S}\frac{D_i}{\pi_i(\hat {\boldsymbol\beta})}\frac{1}{\sigma_i^2+\hat{\tau}_{IPW}^2}} \overset{p}{\to}
\frac{E\{\frac{D_i}{\pi_i(\boldsymbol  \beta_*)}\frac{1}{\sigma_i^2+\tau^2} y_i\}}
{E\{\frac{D_i}{\pi_i(\boldsymbol \beta_*)}\frac{1}{\sigma_i^2+\tau^2} \}}.
\end{align*}  
Noting that $E\{D_i\mid y_i, \sigma_i, n_i\}=\pi_i(\boldsymbol \beta_*)$, the numerator is
\begin{align*}
E\Big\{\frac{D_i}{\pi_i(\boldsymbol\beta_*)}\frac{1}{\sigma_i^2+\tau^2} y_i\Big\} &  =E\Big\{\frac{y_i}{\pi_i(\boldsymbol\beta_*)}\frac{1}{\sigma_i^2+\tau^2}E(D_i\mid y_i, \sigma_i, n_i)\Big\}
= E\Big(\frac{1}{\sigma_i^2+\tau^2}y_i\Big) 
\nonumber \\
&=  E\Big(\frac{1}{\sigma_i^2+\tau^2}(\mu+\sqrt{\sigma_i^2+\tau^2} \epsilon_i)\Big)
=\mu E\Big(\frac{1}{\sigma_i^2+\tau^2}\Big).
\end{align*}
Similarly, the denominator is given by 
\begin{align*}
E\Big\{\frac{D_i}{\pi_i(\boldsymbol \beta_*)}\frac{1}{\sigma_i^2+\tau^2} \Big\}
=E\Big(\frac{1}{\sigma_i^2+\tau^2}\Big),
\end{align*}
and then $\hat{\mu}_{IPW}$ converges in probability to $\mu$. 

\section*{Appendix B: The asymptotic variance with sandwich variance estimator}
Let $\hat{\boldsymbol\theta}^T=(\hat{\boldsymbol\beta}^T,\hat{\tau}_{IPW}^2,\hat{\mu}_{IPW})$, one can see that $\hat{\boldsymbol\theta}$ is asymptotically equivalent to the solution of the following estimating equations  
\begin{align*}
U^{\boldsymbol\theta}(\boldsymbol\theta)&=\sum_{i=1}^S\left(\begin{array}{c}
(1-\frac{D_i}{\pi_i(\boldsymbol\beta)})g(n_i)\\
\frac{1}{\sigma_i^2}\frac{D_i}{\pi_i(\boldsymbol\beta)}\left\{(y_i-\mu)^2-\tau^2\right\}-1\\
\frac{1}{\sigma_i^2+\tau^2}\frac{D_i}{\pi_i(\boldsymbol\beta)}(y_i-\mu)\\
\end{array}
\right)=\boldsymbol0\\
&=\sum_{i=1}^S\left(\begin{array}{c}
U_i^{\boldsymbol\beta}\\
U_i^{\tau^2}\\
U_i^{\mu}\\
\end{array}
\right)=\sum_{i=1}^SU_i^{\boldsymbol\theta}(\boldsymbol\theta).
\end{align*}

Since we have proved the consistency of $\hat{\boldsymbol\theta}$, by applying the theory of $M$-estimation (see Section 2 in the review by Stefanski et al.\cite{stefanski2002}), we could obtain that
\begin{align*}
\sqrt{S}(\hat{\boldsymbol\theta}-\boldsymbol\theta)\simeq-\left\{\frac{1}{S}\sum_{i=1}^S\frac{\partial}{\partial\boldsymbol\theta^T}U_i^{\boldsymbol\theta}(\boldsymbol\theta)\right\}^{-1}\frac{1}{\sqrt{S}}\sum_{i=1}^SU_i^{\boldsymbol\theta}(\boldsymbol\theta).
\end{align*}

This expression entails asymptotic normality of $\sqrt{S}(\hat{\boldsymbol\theta}-\boldsymbol\theta)$ and its variance is consistently estimated by

\begin{align*}
Var\left[\sqrt{S}(\hat{\boldsymbol\theta}-\boldsymbol\theta)\right]\simeq\left\{\frac{1}{S}\sum_{i=1}^S\frac{\partial}{\partial\boldsymbol\theta^T}U_i^{\boldsymbol\theta}(\hat{\boldsymbol\theta})\right\}^{-1}\frac{1}{S}\sum_{i=1}^SU_i^{\boldsymbol\theta}(\hat{\boldsymbol\theta})U_i^{\boldsymbol\theta}(\hat{\boldsymbol\theta})^T
\left\{\frac{1}{S}\sum_{i=1}^S\frac{\partial}{\partial\boldsymbol{\theta}^T}U_i^{\boldsymbol\theta}(\hat{\boldsymbol\theta})\right\}^{-1}
\end{align*}

\section*{Appendix C: Additional simulation studies}
In this appendix, we presented the results of additional simulation studies with {\it sDataset 2} and {\it sDataset 4}. The findings were similar with the results reported in the main text (see Tables 1 to 4). In Tables~\ref{s1} and~\ref{s2}, the simulation results for estimation of $\mu$ and $\tau^2$ for {\it sDataset 2} were presented. We observed that the IPW method with both one-parameter selection functions ( one-parameter logistic (3) and its modified version (4) ) successfully reduced certain biases and misspecification of the selection function can lead certain biases for $\mu$ estimation with large number of studies (S = 50 and 100). Tables~\ref{s3} and~\ref{s4} summarized the simulation results for estimation of $\mu$ and $\tau^2$ for {\it sDataset 4}, we also observed that misspecification of the selection function can introduce certain biases for $\mu$ estimation, and parametric bootstrap confidence intervals seemed more reasonable in contrast to the asymptotic confidence intervals.

\setcounter{table}{0}
\renewcommand{\thetable}{S\arabic{table}}
\begin{landscape}
\begin{table*} 
\centering
\caption{Simulation results for estimation of $\mu$ under one-parameter modified logistic selection model with $\beta=5$ and $\tau$ = 0.05, 0.15 or 0.30 }\label{s1}
\scalebox{0.6}[0.95]{%
\begin{tabular}{*{4}{l}*{16}{c}}
\hline
&&&&\multicolumn{4}{c}{$S=15$}&\multicolumn{4}{c}{$S=25$}&\multicolumn{4}{c}{$S=50$}&\multicolumn{4}{c}{$S=100$}\\
\cmidrule(lr){5-8}\cmidrule(lr){9-12}\cmidrule(lr){13-16}\cmidrule(lr){17-20}
$\tau^2$&Method&Selection&Status&AVE(SD)&CP&LOCI&NOC&AVE(SD)&CP&LOCI&NOC&AVE(SD)&CP&LOCI&NOC&AVE(SD)&CP&LOCI&NOC\\
\hline
0.0025    
   
 & DL &  &  & -0.544 ( 0.081 ) & 0.934 & 0.337 & 1000 & -0.531 ( 0.064 ) & 0.933 & 0.255 & 1000 & -0.530 ( 0.044 ) & 0.902 & 0.174 & 1000 & -0.528 ( 0.031 ) & 0.853 & 0.121 & 1000\\

 & Preston  & 1-logit & M & -0.317 ( 0.738 ) & 0.738 & 14.893 & 809 & -0.401 ( 0.294 ) & 0.719 & 0.247 & 797 & -0.442 ( 0.167 ) & 0.754 & 1.723 & 806 & -0.465 ( 0.066 ) & 0.682 & 6.256 & 804\\

 &  & 1-mlogit & C & -0.453 ( 0.213 ) & 0.805 & 14.686 & 848 & -0.469 ( 0.109 ) & 0.817 & 0.248 & 821 & -0.486 ( 0.052 ) & 0.809 & 0.167 & 839 & -0.489 ( 0.035 ) & 0.824 & 0.119 & 790\\

 & IPW (Asym) & 1-logit & M & -0.509 ( 0.086 ) & 0.866 & 0.276 & 1000 & -0.494 ( 0.068 ) & 0.899 & 0.227 & 1000 & -0.494 ( 0.046 ) & 0.925 & 0.167 & 1000 & -0.492 ( 0.031 ) & 0.936 & 0.120 & 1000\\

 & IPW(Boot) & 1-logit & M & -0.509 ( 0.086 ) & 0.968 & 0.362 & 1000 & -0.494 ( 0.068 ) & 0.955 & 0.271 & 1000 & -0.494 ( 0.046 ) & 0.958 & 0.184 & 1000 & -0.492 ( 0.031 ) & 0.952 & 0.127 & 1000\\

 & IPW (Asym) & 1-mlogit & C & -0.514 ( 0.086 ) & 0.873 & 0.279 & 1000 & -0.499 ( 0.068 ) & 0.902 & 0.229 & 1000 & -0.499 ( 0.047 ) & 0.924 & 0.169 & 1000 & -0.498 ( 0.031 ) & 0.943 & 0.120 & 1000\\

& IPW(Boot) & 1-mlogit & C & -0.514 ( 0.086 ) & 0.971 & 0.381 & 1000 & -0.499 ( 0.068 ) & 0.959 & 0.288 & 1000 & -0.499 ( 0.047 ) & 0.966 & 0.199 & 1000 & -0.498 ( 0.031 ) & 0.966 & 0.137 & 1000\\

\hline
0.0225    
& DL &  &  & -0.549 ( 0.097 ) & 0.915 & 0.376 & 1000 & -0.545 ( 0.077 ) & 0.875 & 0.281 & 1000 & -0.541 ( 0.052 ) & 0.857 & 0.193 & 1000 & -0.538 ( 0.036 ) & 0.813 & 0.136 & 1000\\

 & Preston  & 1-logit & M & -0.268 ( 0.627 ) & 0.689 & 23.957 & 777 & -0.358 ( 0.463 ) & 0.665 & 41.440 & 800 & -0.399 ( 0.239 ) & 0.662 & 15.499 & 779 & -0.426 ( 0.125 ) & 0.632 & 0.157 & 780\\

 &  & 1-mlogit & C & -0.434 ( 0.255 ) & 0.737 & 2.493 & 821 & -0.470 ( 0.119 ) & 0.765 & 0.280 & 844 & -0.481 ( 0.070 ) & 0.789 & 2.986 & 834 & -0.485 ( 0.047 ) & 0.795 & 1.520 & 844\\

 & IPW (Asym) & 1-logit & M & -0.501 ( 0.107 ) & 0.867 & 0.330 & 1000 & -0.498 ( 0.080 ) & 0.881 & 0.264 & 1000 & -0.493 ( 0.055 ) & 0.909 & 0.197 & 1000 & -0.491 ( 0.037 ) & 0.928 & 0.142 & 1000\\

& IPW(Boot) & 1-logit & M & -0.501 ( 0.107 ) & 0.944 & 0.395 & 1000 & -0.498 ( 0.080 ) & 0.919 & 0.293 & 1000 & -0.493 ( 0.055 ) & 0.927 & 0.202 & 1000 & -0.491 ( 0.037 ) & 0.926 & 0.141 & 1000\\

 & IPW (Asym) & 1-mlogit & C & -0.506 ( 0.107 ) & 0.869 & 0.335 & 1000 & -0.504 ( 0.081 ) & 0.885 & 0.268 & 1000 & -0.500 ( 0.056 ) & 0.914 & 0.198 & 1000 & -0.498 ( 0.037 ) & 0.935 & 0.143 & 1000\\

& IPW(Boot) & 1-mlogit & C & -0.506 ( 0.107 ) & 0.950 & 0.417 & 1000 & -0.504 ( 0.081 ) & 0.939 & 0.312 & 1000 & -0.500 ( 0.056 ) & 0.943 & 0.217 & 1000 & -0.498 ( 0.037 ) & 0.946 & 0.151 & 1000\\

\hline
0.0900   
& DL &  &  & -0.583 ( 0.127 ) & 0.855 & 0.468 & 1000 & -0.582 ( 0.096 ) & 0.831 & 0.361 & 1000 & -0.574 ( 0.069 ) & 0.793 & 0.259 & 1000 & -0.577 ( 0.049 ) & 0.620 & 0.184 & 1000\\

 & Preston  & 1-logit & M & -0.252 ( 0.771 ) & 0.643 & 64.287 & 739 & -0.319 ( 0.442 ) & 0.634 & 42.876 & 762 & -0.343 ( 0.250 ) & 0.564 & 29.354 & 732 & -0.360 ( 0.162 ) & 0.503 & 33.319 & 678\\

 & & 1-mlogit & C & -0.429 ( 0.309 ) & 0.692 & 36.410 & 778 & -0.455 ( 0.167 ) & 0.728 & 20.654 & 802 & -0.459 ( 0.119 ) & 0.721 & 9.193 & 784 & -0.472 ( 0.076 ) & 0.739 & 22.365 & 789\\

 & IPW (Asym) & 1-logit & M & -0.510 ( 0.133 ) & 0.856 & 0.427 & 1000 & -0.504 ( 0.100 ) & 0.904 & 0.348 & 1000 & -0.491 ( 0.073 ) & 0.916 & 0.261 & 1000 & -0.492 ( 0.051 ) & 0.922 & 0.189 & 1000\\

& IPW(Boot) & 1-logit & M & -0.510 ( 0.133 ) & 0.912 & 0.483 & 1000 & -0.504 ( 0.100 ) & 0.930 & 0.369 & 1000 & -0.491 ( 0.073 ) & 0.917 & 0.260 & 1000 & -0.492 ( 0.051 ) & 0.912 & 0.183 & 1000\\

 & IPW (Asym) & 1-mlogit & C & -0.516 ( 0.135 ) & 0.853 & 0.551 & 1000 & -0.512 ( 0.101 ) & 0.895 & 0.354 & 1000 & -0.498 ( 0.076 ) & 0.907 & 0.266 & 1000 & -0.499 ( 0.053 ) & 0.924 & 0.193 & 1000\\

 & IPW(Boot) & 1-mlogit & C & -0.516 ( 0.135 ) & 0.917 & 0.510 & 1000 & -0.512 ( 0.101 ) & 0.937 & 0.394 & 1000 & -0.498 ( 0.076 ) & 0.930 & 0.280 & 1000 & -0.499 ( 0.053 ) & 0.929 & 0.197 & 1000\\

\hline
                                         &&&True&-0.500&-&-&-&-0.500&-&-&-&-0.500&-&-&-&-0.500&-&-&-\\
\hline
\end{tabular}}
\flushleft {Selection, the selection model used for estimation: 1-logit denotes the one-parameter logistic selection model, 1-mlogit denotes the one-parameter modified logistic selection model; Status, model specification: C means selection model correctly specified, M means selection model misspecified; s, number of total studies; AVE, mean value of estimates; SD, standard error of estimates; CP, 95\%confidence interval coverage probability; LOCI, length of confidence interval; NOC, number of converged cases; DL, random-effects model with DerSimonian-Laird method; Preston, Preston's conditional likelihood method; IPW (Asym), the proposed method with asymptotic variance; IPW (Boot), the proposed method with parametric bootstrap confidence interval}
\end{table*}
\end{landscape}

\begin{landscape}
\begin{table*}[htbp] 
\centering
\caption{Simulation results for estimation of $\tau^2$ under one-parameter modified logistic selection model with $\beta=5$ and $\tau$ = 0.05, 0.15 or 0.30 }\label{s2}
\scalebox{0.65}[1]{%
\begin{tabular}{*{4}{l}*{16}{c}}
\hline
&&&&\multicolumn{4}{c}{$S=15$}&\multicolumn{4}{c}{$S=25$}&\multicolumn{4}{c}{$S=50$}&\multicolumn{4}{c}{$S=100$}\\
\cmidrule(lr){5-8}\cmidrule(lr){9-12}\cmidrule(lr){13-16}\cmidrule(lr){17-20}
$\tau^2$&Method&Selection&Status&AVE(SD)&CP&LOCI&NOZ&AVE(SD)&CP&LOCI&NOZ&AVE(SD)&CP&LOCI&NOZ&AVE(SD)&CP&LOCI&NOZ\\
\hline
0.0025       

& DL &  &  & 0.009 ( 0.019 ) & 0.952 & 0.230 & 639 & 0.007 ( 0.014 ) & 0.947 & 0.123 & 622 & 0.004 ( 0.008 ) & 0.939 & 0.055 & 644 & 0.002 ( 0.005 ) & 0.902 & 0.029 & 665\\

 & IPW (Asym) & 1-logit & M & 0.010 ( 0.025 ) & 0.994 & 0.061 & 668 & 0.010 ( 0.020 ) & 0.990 & 0.052 & 605 & 0.007 ( 0.013 ) & 0.993 & 0.040 & 574 & 0.005 ( 0.009 ) & 0.997 & 0.030 & 531\\

& IPW(Boot) & 1-logit & M & 0.010 ( 0.025 ) & 0.997 & 0.121 & 668 & 0.010 ( 0.020 ) & 0.999 & 0.088 & 605 & 0.007 ( 0.013 ) & 0.999 & 0.058 & 574 & 0.005 ( 0.009 ) & 0.995 & 0.038 & 531\\

 & IPW (Asym) & 1-mlogit & C &0.010 ( 0.026 ) & 0.993 & 0.062 & 674 & 0.010 ( 0.020 ) & 0.992 & 0.053 & 621 & 0.007 ( 0.014 ) & 0.995 & 0.041 & 602 & 0.005 ( 0.010 ) & 0.996 & 0.031 & 564\\

 & IPW(Boot) & 1-mlogit & C & 0.010 ( 0.026 ) & 0.998 & 0.132 & 674 & 0.010 ( 0.020 ) & 0.999 & 0.101 & 621 & 0.007 ( 0.014 ) & 0.995 & 0.072 & 602 & 0.005 ( 0.010 ) & 0.991 & 0.052 & 564\\

\hline
0.0225   
 & DL &  &  & 0.024 ( 0.034 ) & 0.957 & 0.317 & 409 & 0.019 ( 0.023 ) & 0.941 & 0.165 & 387 & 0.015 ( 0.017 ) & 0.921 & 0.086 & 304 & 0.014 ( 0.013 ) & 0.897 & 0.053 & 196\\

 & IPW (Asym) & 1-logit & M & 0.026 ( 0.044 ) & 0.953 & 0.087 & 430 & 0.023 ( 0.030 ) & 0.958 & 0.074 & 377 & 0.021 ( 0.022 ) & 0.943 & 0.060 & 260 & 0.020 ( 0.017 ) & 0.892 & 0.050 & 152\\

& IPW(Boot) & 1-logit & M & 0.026 ( 0.044 ) & 0.999 & 0.159 & 430 & 0.023 ( 0.030 ) & 0.999 & 0.115 & 377 & 0.021 ( 0.022 ) & 1.000 & 0.081 & 260 & 0.020 ( 0.017 ) & 0.992 & 0.060 & 152\\

 & IPW (Asym) & 1-mlogit &C & 0.026 ( 0.045 ) & 0.958 & 0.089 & 436 & 0.022 ( 0.030 ) & 0.960 & 0.074 & 396 & 0.020 ( 0.022 ) & 0.949 & 0.060 & 281 & 0.019 ( 0.017 ) & 0.895 & 0.049 & 177\\

& IPW(Boot) & 1-mlogit & C & 0.026 ( 0.045 ) & 0.999 & 0.169 & 436 & 0.022 ( 0.030 ) & 0.999 & 0.126 & 396 & 0.020 ( 0.022 ) & 0.999 & 0.094 & 281 & 0.019 ( 0.017 ) & 0.995 & 0.071 & 177\\

\hline
0.0900 
  & DL &  &  & 0.071 ( 0.072 ) & 0.951 & 0.473 & 176 & 0.068 ( 0.057 ) & 0.936 & 0.281 & 80 & 0.068 ( 0.039 ) & 0.908 & 0.170 & 23 & 0.069 ( 0.029 ) & 0.862 & 0.110 & 0\\

 & IPW (Asym) & 1-logit & M &0.075 ( 0.079 ) & 0.665 & 0.164 & 188 & 0.079 ( 0.067 ) & 0.727 & 0.154 & 87 & 0.086 ( 0.050 ) & 0.807 & 0.139 & 16 & 0.090 ( 0.036 ) & 0.861 & 0.115 & 0\\

& IPW(Boot) & 1-logit & M & 0.075 ( 0.079 ) & 0.933 & 0.280 & 188 & 0.079 ( 0.067 ) & 0.907 & 0.234 & 87 & 0.086 ( 0.050 ) & 0.916 & 0.185 & 16 & 0.090 ( 0.036 ) & 0.933 & 0.140 & 0\\

 & IPW (Asym) & 1-mlogit & C & 0.074 ( 0.078 ) & 0.672 & 0.201 & 192 & 0.076 ( 0.065 ) & 0.732 & 0.156 & 96 & 0.081 ( 0.048 ) & 0.789 & 0.139 & 24 & 0.084 ( 0.034 ) & 0.834 & 0.113 & 0\\

 & IPW(Boot) & 1-mlogit & C & 0.074 ( 0.078 ) & 0.939 & 0.289 & 192 & 0.076 ( 0.065 ) & 0.932 & 0.239 & 96 & 0.081 ( 0.048 ) & 0.944 & 0.188 & 24 & 0.084 ( 0.034 ) & 0.949 & 0.142 & 0\\

\hline
\end{tabular}}
\flushleft {Selection, the selection model used for estimation: 1-logit denotes the one-parameter logistic selection model, 1-mlogit denotes the one-parameter modified logistic selection model; Status, model specification: C means selection model correctly specified, M means selection model misspecified; s, number of total studies; AVE, mean value of estimates; SD, standard error of estimates; CP, 95\%confidence interval coverage probability; LOCI, length of confidence interval; NOZ, number of 0 estimates; DL, random-effects model with DerSimonian-Laird method;  IPW (Asym), the proposed method with asymptotic variance; IPW (Boot), the proposed method with parametric bootstrap confidence interval}
\end{table*}
\end{landscape}

\newpage

\begin{landscape}
\begin{table*}[htbp] 
\centering
\caption{Simulation results for estimation of $\mu$ under two-parameter logistic selection model with $\boldsymbol\beta=(-0.3,-1.0)$ and $\tau$ = 0.05, 0.15 or 0.30 }\label{s3}
\scalebox{0.6}[0.95]{%
\begin{tabular}{*{4}{l}*{16}{c}}
\hline
&&&&\multicolumn{4}{c}{$S=15$}&\multicolumn{4}{c}{$S=25$}&\multicolumn{4}{c}{$S=50$}&\multicolumn{4}{c}{$S=100$}\\
\cmidrule(lr){5-8}\cmidrule(lr){9-12}\cmidrule(lr){13-16}\cmidrule(lr){17-20}
$\tau^2$&Method&Selection&Status&AVE(SD)&CP&LOCI&NOC&AVE(SD)&CP&LOCI&NOC&AVE(SD)&CP&LOCI&NOC&AVE(SD)&CP&LOCI&NOC\\
\hline
0.0025       
& DL &  &  & -0.555 ( 0.092 ) & 0.922 & 0.368 & 1000 & -0.546 ( 0.064 ) & 0.926 & 0.272 & 1000 & -0.547 ( 0.046 ) & 0.838 & 0.185 & 1000 & -0.542 ( 0.030 ) & 0.776 & 0.128 & 1000\\

 & Copas & 2-probit & M & -0.503 ( 0.108 ) & 0.501 & 0.197 & 939 & -0.496 ( 0.081 ) & 0.576 & 0.167 & 987 & -0.496 ( 0.055 ) & 0.691 & 0.135 & 998 & -0.492 ( 0.037 ) & 0.736 & 0.102 & 999\\

 & IPW (Asym) & 2-logit & C & -0.508 ( 0.098 ) & 0.901 & 1.350 & 1000 & -0.498 ( 0.071 ) & 0.948 & 2.025 & 1000 & -0.501 ( 0.048 ) & 0.963 & 0.408 & 1000 & -0.498 ( 0.035 ) & 0.979 & 0.269 & 1000\\

& IPW(Boot) & 2-logit & C & -0.508 ( 0.098 ) & 0.957 & 0.398 & 1000 & -0.498 ( 0.071 ) & 0.974 & 0.306 & 1000 & -0.501 ( 0.048 ) & 0.982 & 0.224 & 1000 & -0.498 ( 0.035 ) & 0.984 & 0.167 & 1000\\

 & IPW (Asym) & 2-probit & M & -0.486 ( 0.110 ) & 0.887 & 1.343 & 1000 & -0.468 ( 0.090 ) & 0.942 & 1.630 & 1000 & -0.464 ( 0.070 ) & 0.958 & 5.147 & 1000 & -0.457 ( 0.056 ) & 0.949 & 0.556 & 1000\\

& IPW(Boot) & 2-probit & M & -0.486 ( 0.110 ) & 0.957 & 0.452 & 1000 & -0.468 ( 0.090 ) & 0.969 & 0.369 & 1000 & -0.464 ( 0.070 ) & 0.961 & 0.296 & 1000 & -0.457 ( 0.056 ) & 0.950 & 0.252 & 1000\\

\hline
0.0225     
 & DL &  &  & -0.566 ( 0.105 ) & 0.879 & 0.391 & 1000 & -0.563 ( 0.075 ) & 0.866 & 0.296 & 1000 & -0.564 ( 0.056 ) & 0.763 & 0.204 & 1000 & -0.559 ( 0.037 ) & 0.645 & 0.143 & 1000\\

 & Copas & 2-probit & M & -0.513 ( 0.122 ) & 0.489 & 0.196 & 948 & -0.503 ( 0.100 ) & 0.540 & 0.174 & 992 & -0.502 ( 0.074 ) & 0.578 & 0.134 & 998 & -0.494 ( 0.051 ) & 0.606 & 0.102 & 1000\\

 & IPW (Asym) & 2-logit & C & -0.508 ( 0.110 ) & 0.918 & 1.411 & 1000 & -0.501 ( 0.082 ) & 0.949 & 2.039 & 1000 & -0.503 ( 0.057 ) & 0.953 & 0.659 & 1000 & -0.500 ( 0.041 ) & 0.967 & 0.740 & 1000\\

 & IPW(Boot) & 2-logit & C & -0.508 ( 0.110 ) & 0.948 & 0.420 & 1000 & -0.501 ( 0.082 ) & 0.959 & 0.326 & 1000 & -0.503 ( 0.057 ) & 0.965 & 0.240 & 1000 & -0.500 ( 0.041 ) & 0.974 & 0.182 & 1000\\

 & IPW (Asym) & 2-probit & M & -0.483 ( 0.125 ) & 0.901 & 1.358 & 1000 & -0.467 ( 0.100 ) & 0.934 & 3.655 & 1000 & -0.461 ( 0.078 ) & 0.957 & 0.935 & 1000 & -0.451 ( 0.064 ) & 0.958 & 0.725 & 1000\\

 & IPW(Boot) & 2-probit & M & -0.483 ( 0.125 ) & 0.952 & 0.480 & 1000 & -0.467 ( 0.100 ) & 0.948 & 0.390 & 1000 & -0.461 ( 0.078 ) & 0.955 & 0.311 & 1000 & -0.451 ( 0.065 ) & 0.928 & 0.261 & 1000\\

\hline
0.0900  
 & DL &  &  & -0.612 ( 0.134 ) & 0.830 & 0.486 & 1000 & -0.616 ( 0.099 ) & 0.754 & 0.373 & 1000 & -0.616 ( 0.071 ) & 0.601 & 0.264 & 1000 & -0.615 ( 0.048 ) & 0.331 & 0.188 & 1000\\

 & Copas & 2-probit & M & -0.544 ( 0.168 ) & 0.371 & 0.204 & 961 & -0.545 ( 0.134 ) & 0.389 & 0.172 & 988 & -0.535 ( 0.104 ) & 0.430 & 0.139 & 999 & -0.521 ( 0.083 ) & 0.409 & 0.102 & 1000\\

 & IPW (Asym) & 2-logit & C & -0.519 ( 0.142 ) & 0.912 & 2.492 & 1000 & -0.515 ( 0.108 ) & 0.940 & 1.529 & 1000 & -0.510 ( 0.075 ) & 0.960 & 1.294 & 1000 & -0.510 ( 0.058 ) & 0.981 & 9.322 & 1000\\

& IPW(Boot) & 2-logit & C & -0.519 ( 0.142 ) & 0.919 & 0.501 & 1000 & -0.515 ( 0.108 ) & 0.928 & 0.393 & 1000 & -0.510 ( 0.075 ) & 0.949 & 0.290 & 1000 & -0.510 ( 0.058 ) & 0.952 & 0.224 & 1000\\

 & IPW (Asym) & 2-probit & M & -0.489 ( 0.156 ) & 0.891 & 1.887 & 1000 & -0.471 ( 0.130 ) & 0.940 & 13.522 & 1000 & -0.459 ( 0.094 ) & 0.962 & 6.704 & 1000 & -0.442 ( 0.084 ) & 0.971 & 2.354 & 1000\\

& IPW(Boot) & 2-probit & M & -0.489 ( 0.156 ) & 0.927 & 0.560 & 1000 & -0.471 ( 0.130 ) & 0.936 & 0.460 & 1000 & -0.459 ( 0.094 ) & 0.954 & 0.360 & 1000 & -0.442 ( 0.084 ) & 0.914 & 0.302 & 1000\\

\hline
                                         &&&True&-0.500&-&-&-&-0.500&-&-&-&-0.500&-&-&-&-0.500&-&-&-\\
\hline
\end{tabular}}
\flushleft {Selection, the selection model used for estimation: 2-logit denotes the two-parameter logistic selection model, 2-probit denotes the two-parameter probit selection model; Status, model specification: C means selection model was correctly specified, M means selection model was misspecified; s, number of total studies; AVE, mean value of estimates; SD, standard error of estimates; CP, 95\%confidence interval coverage probability; LOCI, length of confidence interval; NOC, number of converged cases; DL, random-effects model with DerSimonian-Laird method; Copas, Copas' sensitivity analysis method; IPW (Asym), the proposed method with asymptotic variance; IPW (Boot), the proposed method with parametric bootstrap confidence interval}
\end{table*}
\end{landscape}

\begin{landscape}
\begin{table*}[htbp] 
\centering
\caption{Simulation results for estimation of $\tau^2$ under two-parameter logistic selection model with $\boldsymbol\beta=(-0.3,-1.0)$ and $\tau$ = 0.05, 0.15 or 0.30 }\label{s4}
\scalebox{0.65}[1]{%
\begin{tabular}{*{4}{l}*{16}{c}}
\hline
&&&&\multicolumn{4}{c}{$S=15$}&\multicolumn{4}{c}{$S=25$}&\multicolumn{4}{c}{$S=50$}&\multicolumn{4}{c}{$S=100$}\\
\cmidrule(lr){5-8}\cmidrule(lr){9-12}\cmidrule(lr){13-16}\cmidrule(lr){17-20}
$\tau^2$&Method&Selection&Status&AVE(SD)&CP&LOCI&NOZ&AVE(SD)&CP&LOCI&NOZ&AVE(SD)&CP&LOCI&NOZ&AVE(SD)&CP&LOCI&NOZ\\
\hline
0.0025       
 & DL &  &  & 0.010 ( 0.022 ) & 0.960 & 0.325 & 644 & 0.007 ( 0.015 ) & 0.955 & 0.150 & 642 & 0.004 ( 0.009 ) & 0.937 & 0.063 & 684 & 0.002 ( 0.005 ) & 0.907 & 0.031 & 713\\

 & IPW (Asym) & 2-logit & C & 0.013 ( 0.033 ) & 0.994 & 0.169 & 671 & 0.012 ( 0.027 ) & 0.998 & 0.207 & 613 & 0.008 ( 0.016 ) & 0.995 & 0.075 & 599 & 0.006 ( 0.014 ) & 0.996 & 0.055 & 596\\

& IPW(Boot) & 2-logit & C & 0.013 ( 0.033 ) & 0.988 & 0.143 & 671 & 0.012 ( 0.027 ) & 0.997 & 0.109 & 613 & 0.008 ( 0.016 ) & 0.998 & 0.080 & 599 & 0.006 ( 0.014 ) & 0.990 & 0.061 & 597\\

 & IPW (Asym) & 2-probit & M & 0.017 ( 0.040 ) & 0.991 & 0.156 & 635 & 0.020 ( 0.043 ) & 0.984 & 0.245 & 519 & 0.020 ( 0.037 ) & 0.970 & 0.537 & 428 & 0.021 ( 0.033 ) & 0.971 & 0.133 & 305\\

& IPW(Boot) & 2-probit & M & 0.017 ( 0.040 ) & 0.992 & 0.179 & 635 & 0.020 ( 0.043 ) & 0.985 & 0.154 & 518 & 0.020 ( 0.037 ) & 0.949 & 0.130 & 428 & 0.021 ( 0.033 ) & 0.921 & 0.117 & 304\\

\hline
0.0225 
  & DL &  &  & 0.018 ( 0.030 ) & 0.964 & 0.357 & 524 & 0.017 ( 0.025 ) & 0.951 & 0.192 & 412 & 0.013 ( 0.017 ) & 0.923 & 0.092 & 363 & 0.012 ( 0.012 ) & 0.899 & 0.055 & 247\\

 & IPW (Asym) & 2-logit & C & 0.020 ( 0.038 ) & 0.968 & 0.206 & 551 & 0.022 ( 0.035 ) & 0.972 & 0.294 & 403 & 0.020 ( 0.025 ) & 0.966 & 0.129 & 311 & 0.019 ( 0.019 ) & 0.933 & 0.145 & 190\\

& IPW(Boot) & 2-logit & C & 0.020 ( 0.038 ) & 0.998 & 0.166 & 551 & 0.022 ( 0.035 ) & 1.000 & 0.133 & 403 & 0.020 ( 0.025 ) & 0.996 & 0.101 & 311 & 0.019 ( 0.019 ) & 0.993 & 0.079 & 190\\

 & IPW (Asym) & 2-probit & M & 0.025 ( 0.048 ) & 0.968 & 0.195 & 511 & 0.033 ( 0.049 ) & 0.964 & 0.797 & 345 & 0.035 ( 0.043 ) & 0.960 & 0.191 & 196 & 0.040 ( 0.038 ) & 0.957 & 0.175 & 72\\

 & IPW(Boot) & 2-probit & M & 0.025 ( 0.048 ) & 0.995 & 0.204 & 511 & 0.033 ( 0.049 ) & 0.993 & 0.179 & 345 & 0.035 ( 0.043 ) & 0.964 & 0.151 & 199 & 0.040 ( 0.038 ) & 0.910 & 0.131 & 71\\

\hline
0.0900 
 & DL &  &  & 0.059 ( 0.066 ) & 0.941 & 0.598 & 253 & 0.058 ( 0.052 ) & 0.923 & 0.318 & 128 & 0.054 ( 0.037 ) & 0.880 & 0.173 & 42 & 0.055 ( 0.026 ) & 0.790 & 0.111 & 9\\

 & IPW (Asym) & 2-logit & C & 0.063 ( 0.073 ) & 0.677 & 0.336 & 271 & 0.069 ( 0.065 ) & 0.730 & 0.292 & 128 & 0.071 ( 0.047 ) & 0.806 & 0.297 & 38 & 0.074 ( 0.036 ) & 0.837 & 2.051 & 3\\

& IPW(Boot) & 2-logit & C & 0.063 ( 0.073 ) & 0.937 & 0.276 & 271 & 0.069 ( 0.065 ) & 0.932 & 0.236 & 128 & 0.071 ( 0.047 ) & 0.938 & 0.185 & 38 & 0.074 ( 0.036 ) & 0.943 & 0.149 & 3\\

 & IPW (Asym) & 2-probit & M & 0.071 ( 0.083 ) & 0.688 & 0.383 & 250 & 0.084 ( 0.080 ) & 0.769 & 2.388 & 106 & 0.092 ( 0.060 ) & 0.875 & 1.039 & 26 & 0.105 ( 0.056 ) & 0.916 & 0.586 & 2\\

& IPW(Boot) & 2-probit & M & 0.071 ( 0.083 ) & 0.956 & 0.315 & 250 & 0.084 ( 0.080 ) & 0.963 & 0.282 & 106 & 0.092 ( 0.060 ) & 0.979 & 0.226 & 26 & 0.105 ( 0.056 ) & 0.921 & 0.185 & 2\\

\hline
\end{tabular}}
\flushleft {Selection, the selection model used for estimation: 2-logit denotes the two-parameter logistic selection model, 2-probit denotes the two-parameter probit selection model; Status, model specification: C means selection model correctly specified, M means selection model misspecified; s, number of total studies; AVE, mean value of estimates; SD, standard error of estimates; CP, 95\%confidence interval coverage probability; LOCI, length of confidence interval; NOZ, number of 0 estimates; DL, random-effects model with DerSimonian-Laird method;  IPW (Asym), the proposed method with asymptotic variance; IPW (Boot), the proposed method with parametric bootstrap confidence interval}
\end{table*}
\end{landscape}

\section*{Appendix D: Dataset of Clopidogrel study}

\begin{table}[h]
\centering
\caption{Clopidogrel dataset}
\begin{tabular}{rlrrrrrrrrl}
  \toprule
  &&\multicolumn{2}{c}{High Dose}&\multicolumn{2}{c}{Standard Dose}\\
  \cmidrule(lr){3-4}\cmidrule(lr){5-6}
  
 No.& Study &Events & Total & Events & Total & $n_i$ &$ logOR_i$ &$ \sigma_i $& $D_i$ \\ 
  \midrule
1 & Aradi 2012 & 1 & 36 & 8 & 38 & 74 & -2.23 & 1.09 & 1  \\ 
  2 & DOUBLE 2010 & 0 & 24 & 1 & 24 & 48 & -1.14 & 1.66 & 1  \\ 
  3 & EFFICIENT 2011 & 2 & 47 & 8 & 47 & 94 & -1.53 & 0.82 & 1  \\ 
  4 & GRAVITAS 2011 & 25 & 1109 & 25 & 1105 & 2214 & -0.00 & 0.29 & 1 \\ 
  5 & Gremmel 2011 & 1 & 21 & 2 & 23 & 44 & -0.64 & 1.26 & 1  \\ 
  6 & Han 2009 & 4 & 403 & 9 & 410 & 813 & -0.81 & 0.61 & 1\\ 
  7 & Ren LH 2012 & 6 & 46 & 10 & 55 & 101 & -0.39 & 0.56 & 1  \\ 
  8 & Roghani 2011 & 4 & 205 & 2 & 195 & 400 & 0.65 & 0.87 & 1  \\ 
  9 & Tousek 2011 & 1 & 30 & 2 & 30 & 60 & -0.73 & 1.25 & 1  \\ 
  10 & VASP-02 2008 & 0 & 58 & 1 & 62 & 120 & -1.05 & 1.64 & 1  \\ 
  11 & von Beckerath 2007 & 1 & 31 & 1 & 29 & 60 & -0.07 & 1.44 & 1  \\ 
  12 & Wang 2011 & 14 & 150 & 30 & 156 & 306 & -0.84 & 0.35 & 1  \\ 
  13 & NCT01069302 &  &  &  &  & 106 &  &  & 0  \\ 
  14 & NCT01371058 &  &  &  &  & 350 &  &  & 0  \\ 
  15 & NCT01102439 &  &  &  &  & 82 &  &  & 0  \\
  \bottomrule
\end{tabular}
\end{table}

\end{document}